\documentstyle[aps,preprint,floats,epsf]{revtex}

\tighten
\newcommand{\mpt}{\not{\hbox{\kern-4pt $p_T$}}}
\newcommand{\met}{\not{\hbox{\kern-4pt $E_T$}}}
\newcommand{\fdsb}{F_{\text{DSB}}}
\newcommand{\nn}{N_5}
\newcommand{\mmess}{M}
\newcommand{\lmess}{\Lambda}
\newcommand{\signmu}{\text{sign}(\mu)}
\newcommand{\mmssm}{M_{\text{MSSM}}}
\newcommand{\stau}{\tilde{\tau}}
\newcommand{\stauone}{\tilde{\tau}_1}
\newcommand{\slepton}{\tilde{l}}
\newcommand{\selectron}{\tilde{e}}
\newcommand{\smuon}{\tilde{\mu}}
\newcommand{\snu}{\tilde{\nu}}

\newcommand{\ptcut}{p_T^{\text{cut}}}
\newcommand{\fb}{\text{ fb}}
\newcommand{\ifb}{\text{ fb}^{-1}}
\newcommand{\ipb}{\text{ pb}^{-1}}
\newcommand{\gev}{\text{ GeV}}
\newcommand{\tev}{\text{ TeV}}

\begin{document}

\draft
\pagestyle{empty}

\preprint{
\noindent
\hfill
\begin{minipage}[t]{3in}
\begin{flushright}
LBNL--41133 \\
UCB--PTH--97/60\\
hep-ph/9712499\\
December 1997
\end{flushright}
\end{minipage}
}

\title{
\vspace*{0.4in}
Tevatron Signatures of Long-lived Charged Sleptons\\
in Gauge-Mediated Supersymmetry Breaking Models
\thanks{This work was supported in part by the Director, Office of
Energy Research, Office of High Energy and Nuclear Physics, Division
of High Energy Physics of the U.S. Department of Energy under Contract
DE--AC03--76SF00098, and in part by the National Science Foundation
under grant PHY--95--14797. } }
\author{Jonathan L. Feng and Takeo Moroi
\vspace*{0.2in}
}

\address{Theoretical Physics Group \\
Ernest Orlando Lawrence Berkeley National Laboratory\\
and\\
Department of Physics \\
University of California, Berkeley, California 94720 }

\maketitle

\begin{abstract}
In supersymmetric models with gauge-mediated supersymmetry breaking,
charged sleptons are the next lightest supersymmetric particles and
decay outside the detector for large regions of parameter space.  In
such scenarios, supersymmetry may be discovered by searches for a
number of novel signals, including highly ionizing tracks from
long-lived slow charged particles and excesses of multi-lepton
signals.  We consider this scenario in detail and find that the
currently available Tevatron data probes regions of parameter space
beyond the kinematic reach of LEP II.  Future Tevatron runs with
integrated luminosities of 2, 10, and $30\ifb$ probe right-handed
slepton masses of 110, 180, and 230 GeV and Wino masses of 310, 370,
and 420 GeV, respectively, greatly extending current search limits.
\end{abstract}

\pagestyle{plain}

\section{Introduction}
\label{sec:intro}

Supersymmetry (SUSY) provides an interesting framework for stabilizing
the electroweak scale against radiative corrections. A realistic
realization of this idea is the minimal supersymmetric standard model
(MSSM), the model with minimal field content in which all the standard
model particles have their superpartners.  In the MSSM, however, the
mechanism for SUSY breaking is not specified, but is simply
parametrized by introducing a large number of soft SUSY breaking
parameters by hand.  The MSSM is therefore best viewed as the low
energy effective theory of some more fundamental theory, which must
give a phenomenologically viable explanation of the origin of SUSY
breaking.

Models with gauge-mediated SUSY breaking~\cite{GM,LEGM} are candidates
for such a theory, and provide an appealing way to generate SUSY
breaking soft terms in the MSSM Lagrangian.  In these models, SUSY
breaking originates in a dynamical SUSY breaking sector and is then
mediated via gauge interactions to the MSSM sector. The SUSY breaking
mechanism does not distinguish between flavors, and so, in the MSSM
sector, all scalars with identical gauge quantum numbers have the same
SUSY breaking mass (at the scale where the SUSY breaking fields are
integrated out). Thus, the universality of squark and slepton masses
at some scale is guaranteed, and serious SUSY flavor-changing neutral
current (FCNC) problems can be evaded.

To preserve the natural suppression of FCNCs in gauge-mediated SUSY
breaking models, it is essential that no large intergenerational
scalar mixing terms be introduced by supergravity contributions. In
the absence of assumptions concerning the flavor structure of such
supergravity contributions, these mixing terms are expected to be of
order the gravitino mass $m_{3/2} \sim \fdsb/M_P$, where $\fdsb^{1/2}$
is the scale of dynamical supersymmetry breaking, and $M_P$ is the
Planck mass. FCNC constraints then require $m_{3/2} \ll m_W$.  The
gravitino is therefore the lightest supersymmetric
particle,\footnote{We assume that there are no other exotic light
particles.} which has a number of important implications for collider
phenomenology.  First, the lightest standard model superpartner is now
the next lightest supersymmetric particle (NLSP) and eventually decays
into its standard model partner and the gravitino. The NLSP may then
be charged or colored, as astrophysical and other constraints on such
particles are thereby evaded.  Second, the NLSP may be stable or
unstable in collider experiments, depending on $\fdsb$.  If $\fdsb$ is
high enough, decays to the gravitino is suppressed, and the NLSP
decays outside the detector.

The possible collider phenomena in gauge-mediated models are largely
determined by the character of the NLSP. Typically, in these models
the NLSP is either the lightest neutralino or the stau mass eigenstate
$\stauone$.\footnote{In the latter case, the role of the NLSP may be
shared by the sleptons $\selectron_R$ and $\smuon_R$ if they are
sufficiently degenerate with $\stauone$.}  As this particle may be
either long- or short-lived,\footnote{Throughout this paper,
``long-lived'' and ``short-lived'' refer to particles that typically
decay outside and inside collider experiments, respectively.} there
are four broad possibilities.  If the NLSP is a long-lived neutralino,
the SUSY signatures are the conventional $\met$ signals and are
identical to those present in gravity-mediated models with $R$-parity
conservation.  If the NLSP is a short-lived neutralino, superparticle
production typically results in hard photons, a signal that has also
been studied in great detail~\cite{hardphoton,DTW}.  On the other
hand, the NLSP could be the stau.  In the case of a short-lived stau
NLSP, there are a number of interesting new signatures, which have
been studied in Refs.~\cite{Dutta,longLEP}.

Here, we explore the remaining possible scenario, in which the NLSP is
a long-lived stau.  This scenario is in fact realized in large
portions of the parameter space: as we will discuss in
Sec.~\ref{sec:model}, the stau is often the NLSP in models with
non-minimal messenger sectors, and it is stable for $\fdsb^{1/2} \agt
10^7\gev$, as is also the case in many models, including, for example,
the gauge mediation models of Ref.~\cite{LEGM}.  In this scenario, the
familiar $\met$ and hard photon signals are absent.  However, SUSY may
manifest itself at colliders through a number of spectacular new
phenomena. For example, staus, as massive charged particles, may be
slow and appear as highly ionizing
tracks~\cite{DT,DTW,longLEP,AKM}. If, on the other hand, the staus are
quite relativistic, they cannot be distinguished from muons, and we
may have excesses of dimuon or multi-lepton events as a result of
superparticle production.

In this paper, we present a detailed discussion of the prospects for
SUSY discovery in this scenario, and estimate the discovery reach for
a variety of signatures.  At present, the most stringent bound on
these scenarios comes from searches for stable charged particles at
LEP~\cite{LEP}.  Such searches, using combined data from runs with
center of mass energies up to $\sqrt{s} = 172\gev$, yield the limit
$m_{\stau_R} > 75\gev$~\cite{LEPcombined}, where no degeneracy between
generations is assumed.  In this study, we begin by considering
possible probes from Tevatron Run I data with $\sqrt{s}= 1.8\tev$ and
an integrated luminosity of $L = 110 \ipb$. Working within a specific,
well-motivated model, we find that this currently available data
probes regions of parameter space beyond the current LEP bound and
even beyond the final LEP II kinematic limit.  We then consider
possible improvements at Tevatron Run II and the possible TeV33
upgrade.  As a number of signals are essentially background-free, the
prospects for improvement are especially bright.  For $\sqrt{s} =
2\tev$, and integrated luminosities of $L=2$, 10, and $30\ifb$, we
find that SUSY may be discovered for right-handed slepton masses of
110, 180, and 230 GeV and Wino masses of 310, 370, and 420 GeV,
respectively. While exact discovery reaches must await detailed
experimental analyses beyond the scope of this study, these results
indicate that the prospects for such searches are promising, as future
Tevatron runs will extend current search boundaries far into ranges
typically expected for superpartner masses.

The organization of this paper is as follows. In Sec.~\ref{sec:model},
we briefly review the framework of gauge-mediated supersymmetric
models. We describe a particular, well-motivated model that displays
many of the generic features of gauge mediation models, and which we
will use throughout the rest of the paper.  In the subsequent
sections, we explore three new signatures for long-lived staus.  For
each, we first estimate the discovery reach in terms of physical
slepton and gaugino masses. In Sec.~\ref{sec:ionization}, we discuss
the possibility of detecting the highly ionizing tracks of heavy
charged NLSPs. In Sec.~\ref{sec:dimuon}, we consider a possible excess
of dimuon-like events due to the pair production of $\tilde{\tau}_1$.
The dimuon excess is a relatively weak probe at the Tevatron; we
discuss also its prospects at the LHC.  We then discuss the
multi-lepton signal in Sec.~\ref{sec:multi-lepton}. Finally, in
Sec.~\ref{sec:comparison} we summarize all of these results by
studying and comparing the discovery reaches of the various signals in
the context of the model we consider and its fundamental parameter
space. We present our conclusions in Sec.~\ref{sec:conclusions}.

\section{Gauge-Mediated Models and the Long-lived Stau NLSP}
\label{sec:model}

The probes discussed in the following sections have relevance for
many models, both supersymmetric and non-supersymmetric, with charged
particles that are stable within the detector, and thus they have wide
applicability.  Even within the framework of SUSY, such particles are
possible in a variety of settings.  However, among SUSY scenarios with
long-lived charged particles, by far the most natural are models with
gauge-mediated SUSY breaking, where the NLSP may be charged and decays
to the gravitino with macroscopic decay lengths for a large range of
parameters.\footnote{Long-lived charged particles are also possible in
gravity-mediated models if the decays of a charged superparticle, such
as a chargino or charged slepton, are, for example, highly phase space
suppressed, or are possible only through small $R$-parity violating
couplings.} In this section, we will discuss these gauge-mediated SUSY
breaking scenarios, and determine what parameter ranges lead to the
long-lived charged particle scenario.

In the general framework of gauge mediation, any of the MSSM
superpartners could in principle be the NLSP and stable in collider
detectors.  For concreteness, and to obtain quantitative results, we
will specify a particular, well-motivated model.  In this section, we
briefly review this simple and calculable model, which we will use to
determine the promise of various charged NLSP signatures.  In doing
so, we define our conventions and notation, and introduce the
parameters that will play a central role in the following discussions.
Further details of this model may be found in Refs.~\cite{DTW,BMPZ}.

In the class of gauge mediation models we consider, there are three
sectors: the dynamical SUSY breaking (DSB) sector, the messenger
sector, and the MSSM sector. In the first, the $F$ component of some
chiral superfield condenses, generating a vacuum expectation value
$\fdsb \neq 0$.  $\fdsb$ determines both the mass of the gravitino

\begin{equation}
m_{3/2}=\frac{\fdsb}{\sqrt{3} M_*} \ ,
\end{equation}
where $M_* = M_P/\sqrt{8\pi} \simeq 2.4 \times 10^{18} \gev$ is the
reduced Planck mass, and the strength of its interactions through, for
example, 

\begin{equation}
  {\cal L}_{\rm int} = \frac{m_{\tilde{\tau}_R}^2}{\fdsb}
  \bar{\psi} \tau_R \stau_R^* + \text{h.c.} \ , 
  \label{L_int}
\end{equation}
where $\psi$ is the longitudinal component of the gravitino, the
Goldstino.

This dynamical SUSY breaking is mediated by two-loop effects to the
messenger sector, which contains a singlet field $S$, and, we will
assume, $\nn$ pairs of vector-like ${\bf 5}+{\bf \overline{5}}$
representations of SU(5), that is, $\nn$ vector-like representations
of SU(2)$_L$ doublets, and $\nn$ vector-like representations of
SU(3)$_C$ triplets, where SU(2)$_L$ and SU(3)$_C$ are the standard
model gauge groups. (Such messenger field content preserves gauge
coupling unification.  Vector-like representations ${\bf 10}+{\bf
\overline{10}}$ of SU(5) may also be present, and each is equivalent 
to 3 vector-like ${\bf 5}+{\bf \overline{5}}$ representations for the
following discussion.) The scalar and auxiliary components of the
singlet field then get vacuum expectation values $\langle S \rangle$
and $F_S$, respectively, which generate SUSY breaking masses for the
$\nn$ vector-like fields.  We denote the scale of these masses, the
messenger scale, by $\mmess$.

Finally, once the messenger fields are integrated out at the scale
$\mmess$, soft SUSY breaking masses for the superparticles in the MSSM
sector are generated.  Letting

\begin{eqnarray} 
\lmess\equiv\frac{F_S} {\langle S\rangle} \ ,
\end{eqnarray} 
and assuming $\lmess \ll M$, we find that one-loop diagrams induce
gaugino masses

\begin{eqnarray}
M_i(\mmess) = \nn \, \lmess\, c_i \frac{g_i^2(\mmess)}{16\pi^2} \, ,
\label{m_gaugino}
\end{eqnarray} 
where $i=1,2,3$ for the U(1)$_Y$, SU(2)$_L$, and SU(3)$_C$ groups,
$c_1=\frac{5}{3}$, and $c_2=c_3=1$.  Two-loop diagrams induce the soft
SUSY breaking sfermion squared masses

\begin{eqnarray}
  M_{\tilde{f}}^2(\mmess) = 2\nn \, \lmess^2 \sum_{i=1}^{3}
  C^f_i \left[\frac{g_i^2(\mmess)}{16\pi^2}\right]^{2} \ .
\label{m_sf}
\end{eqnarray}
Here, $C^f_1=\frac{5}{3}Y^2$ with $Y=Q-T_3$ being the usual
hypercharge, and $C^f_i=0$ for gauge singlets, $\frac{3}{4}$ for
SU(2)$_L$ doublets, and $\frac{4}{3}$ for SU(3)$_C$ triplets. Finally,
trilinear scalar couplings ($A$ terms) are also induced, but are
highly suppressed, as they are generated at two-loops, but have
dimensions of mass, not mass squared.  They may be taken to vanish at
the scale $M$.

Once the boundary conditions are given at the messenger scale
$\mmess$, SUSY breaking parameters at the electroweak scale can be
obtained by renormalization group (RG) evolution.\footnote{Throughout
this study, we use one-loop RG equations unless otherwise noted.}  In
particular, the Higgs mass squared can be driven negative by the large
top Yukawa coupling constant, and hence break electroweak symmetry. In
our analysis, we do not specify the mechanism by which the
supersymmetric Higgs mass $\mu$ and SUSY breaking Higgs mixing mass
$m_3^2$ are generated.  Several attempts to explain the origin of
these parameters may be found in the literature~\cite{LEGM,mu}. Here
we regard them as free parameters.  We then fix one combination of
them by demanding the correct value of the Fermi constant using the
tree-level Higgs potential. The remaining freedom may be specified by
choosing $\tan\beta$, the ratio of the vacuum expectation values of
the two neutral Higgs scalars, and $\signmu$, where we follow the
conventions of Ref.~\cite{HK} in defining the sign of $\mu$.  Thus,
the simple model we have defined has 4+1 free parameters:\footnote{If
all the couplings in the model are ${\cal O}(1)$, $\langle S\rangle$
and $F_S^{1/2}$ (and hence $\mmess$ and $\lmess$) are of the same
order. However, in general, there may be a hierarchy between them. For
example, if the coupling constant for the $S^3$ term in the
superpotential is small, and all other couplings in the superpotential
are ${\cal O} (1)$, $\langle S\rangle$ is enhanced relative to $F_S$,
resulting in $\lmess \ll \mmess$~\cite{LEGM}. Thus, in our analysis,
we treat $\mmess$ and $\lmess$ as independent parameters.}

\begin{equation}
\nn\, ,\ \mmess\, ,\ \lmess\, ,\ \tan\beta\, ,\ \signmu \ .
\end{equation}
In this model with the radiative electroweak symmetry breaking
condition, the mass eigenvalues and mixing parameters for all
superparticles are determined once we fix these parameters.

While the determination of SUSY parameters at the electroweak scale
requires detailed calculation, it is useful to discuss the qualitative
features of the superparticle spectrum.  In the model described above,
usually $\mu > M_1, M_2$.  Furthermore, as can be seen in
Eq.~(\ref{m_gaugino}), gaugino masses are proportional to the
corresponding gauge coupling constants, and hence $M_1 < M_2$.  As a
result, $\chi^0_1 \approx \tilde{B}$ with mass $m_{\chi^0_1} \approx
M_1$, and $\chi^0_2 \approx \tilde{W}^3$ and $\chi^\pm_1 \approx
\tilde{W}^\pm$ with masses $m_{\chi^0_2}, m_{\chi^\pm_1} \approx M_2$.
In the scalar sector, sfermion masses are primarily proportional to
the relevant gauge coupling constants. Therefore, squarks are
typically much heavier than sleptons, and right-handed sleptons, which
have only U(1)$_Y$ quantum numbers, are the lightest sfermions.  The
sleptons $\selectron_R$ and $\smuon_R$ are almost degenerate, as
Yukawa coupling effects are negligibly small. On the contrary, the tau
Yukawa coupling may significantly lower the lighter stau mass through
both RG evolution and left-right mixing.  These effects are enhanced
for large $\tan\beta$.  The lightest scalar is thus the lighter stau
$\stauone$, which is predominantly $\stau_R$.

We are interested in the case where $m_{\stauone} < m_{\chi^0_1}$ and
$\stauone$ is the NLSP. This relation is most strongly and obviously
dependent on $\nn$.  As one can see in Eqs.~(\ref{m_gaugino}) and
(\ref{m_sf}), at the messenger scale, $M_i \propto \nn$, while
$M_{\tilde{f}} \propto \sqrt{\nn}$; thus, large $\nn$ reduces
$m_{\stauone}/m_{\chi^0_1}$.  There is also an important dependence on
$M$, as RG evolution increases right-handed slepton masses relative to
gaugino masses, and so large values of $\mmess$ increase
$m_{\stauone}/m_{\chi^0_1}$.  As noted above, large values of
$\tan\beta$ decrease $m_{\stauone}$ and so reduce
$m_{\stauone}/m_{\chi^0_1}$.  Finally, the ratio $m_{\stauone} /
m_{\chi^0_1}$ is independent of $\lmess$ and $\signmu$ to the extent
that the relation $m_{\stauone} / m_{\chi^0_1} \approx M_{\stau_R} /
M_1$ is valid, where $M_{\stau_R}$ and $M_1$ are the soft SUSY
breaking parameters.  

In Fig.~\ref{fig:MmessN5}, we show contours of $M_{\stau_R}(\mmssm) =
M_1(\mmssm)$ in the $(\mmess, \nn)$ plane for various $\tan\beta$ and
$\mmssm = 1 \tev$.  (See also Ref.~\cite{DDRG}.)  The shaded region is
excluded by the requirement that the gauge coupling constants remain
perturbative up to the GUT scale $M_{\text{GUT}} \simeq 2.0\times
10^{16}\gev$ under two-loop RG evolution.  It is important to note
that $M_{\stau_R}$ and $M_1$ are only approximately the physical
$\stauone$ and $\chi^0_1$ masses because of mixing effects and
$D$-term contributions, and, in particular, the difference between
$M_{\stau_R}$ and $m_{\stauone}$ may be large for high $\tan\beta$.
However, this figure gives a qualitative picture of the parameter
region where $\stauone$ is the NLSP.  For a minimal messenger sector
with $\nn=1$, $\stauone$ is not the NLSP (unless $\tan\beta$ is
large~\cite{DTW,BMPZ}).  However, for $\nn \geq 2$, $\stauone$ is the
NLSP for a wide range of parameter space.

In Figs.~\ref{fig:srM1} and \ref{fig:slM2}, we give contour plots of
the soft SUSY breaking parameter ratios $M_{\stau_R}/M_1$ and
$M_{\slepton_L}/M_2$, respectively, where these ratios are at the
scale $\mmssm = 1\tev$.  We fix $\tan\beta = 3$ for these plots.  In
the stau NLSP region allowed by perturbativity, $M_{\stau_R}/M_1 >
0.6$, and for a significant portion of this region, the NLSP and
$\chi^0_1$ are fairly degenerate.  We see also that, in this region,
it is almost always true that $M_{\slepton_L} < M_2$.  In addition,
$M_{\slepton_L}/M_2 > 0.7$, and for a significant portion of this
region, the $\slepton_L$ and Wino states are also fairly degenerate.
These observations will be of use in the following sections.

Given the interaction of Eq.~(\ref{L_int}), the $\stauone$ decay width
is

\begin{eqnarray}
\Gamma_{\stauone}= \frac{1}{16\pi} \frac{m_{\stauone}^5}{\fdsb^2} \ ,
\end{eqnarray} 
and its decay length is

\begin{eqnarray}
  L\simeq 10 \text{ km} \times \langle\beta\gamma\rangle
  \left[\frac{\fdsb^{1/2}}{10^7 \gev}\right]^{4} \left[\frac{100
  \gev}{m_{\stauone}}\right]^5,
\end{eqnarray}
where $\beta$ is the $\stauone$ velocity and $\gamma \equiv
(1-\beta^2)^{-1/2}$. For $\fdsb^{1/2} \agt 10^7\gev$, the NLSP may be
considered stable in detectors, as the likelihood of an NLSP decaying
within the detector is negligible for Tevatron luminosities and the
typical allowed ranges of supersymmetric cross sections. Such values
of $\fdsb^{1/2}$ are expected in many proposed models.  For example,
in typical messenger sector models, $\fdsb^{1/2}$ is likely to be
larger than $\sim 10^7\gev$ if there is no fine-tuning; otherwise, the
SUSY breaking scale in the MSSM sector is so low that slepton masses
become lighter than the experimental limit~\cite{dGMM}.  There are
also many other recently proposed models, including those based on
direct gauge mediation, that give high SUSY breaking
scales\cite{DDRG,HEGM}.\footnote{Note, however, that in a class of
models without messenger sectors~\cite{mgrav=1keV}, $\fdsb^{1/2}$ can
be as small as $10^4$ to $10^5$ GeV, and the NLSP may decay inside the
detector.} In a wide variety of models, therefore, the stau decays
outside the detector, and we will concentrate on this possibility for
the rest of this study.

\section{Highly Ionizing Tracks} 
\label{sec:ionization}

Long-lived staus are charged, weakly interacting, penetrating
particles, and so appear in the tracking and muon chambers of collider
detectors with little associated energy deposition in calorimeters.
They therefore look like muons.  However, since long-lived staus must
have mass above 75 GeV~\cite{LEPcombined}, their velocity is typically
not large if produced at the Tevatron.  Moderately relativistic
charged particles lose energy in matter primarily through
ionization. The rate of energy loss $-dE/dx$ is determined largely by
the particle's velocity $\beta$, grows rapidly with decreasing $\beta$
for low $\beta$~\cite{PDG}, and may be measured in tracking chambers.
As a result, the cross section of slow staus may be significant, and
highly ionizing tracks may be a spectacular signal of such particles.

If the stau is the NLSP, all superparticles eventually decay to staus,
and so all superpartner production processes may result in highly
ionizing tracks.  However, in typical gauge-mediated scenarios,
superparticle production at the Tevatron is dominated by only a few
processes.  First, squarks, Higgsinos, and heavy Higgs bosons (which
may decay to superparticles) are too heavy to be produced in
significant numbers.  Second, left-handed sleptons, though typically
much lighter than these particles, are still significantly heavier
than right-handed sleptons, and so, in terms of overall rate, the
Drell-Yan production of left-handed sleptons is negligible relative to
that for right-handed sleptons. The dominant processes are therefore
right-handed slepton production and gaugino production.  In the
following subsections, we discuss these two processes and their
potential for SUSY discovery in turn.

\subsection{Slepton Production}
\label{sec:slepton}

At the Tevatron, direct right-handed slepton production takes place
through the Drell-Yan processes

\begin{equation}
p\bar{p} \to \gamma^*, Z^* \to \stauone \stauone^* ,\,
\selectron_R \selectron^*_R ,\, \smuon_R \smuon^*_R \ .
\label{slepton}
\end{equation}
Cross sections for a single generation of right-handed slepton
production as a function of slepton mass are given in
Fig.~\ref{fig:sl_prod} for $\sqrt{s} = 1.8 \tev$ and $\sqrt{s} =
2\tev$.\footnote{Next-to-leading order effects have been shown
recently to enhance the tree-level cross sections by 35\% to
40\%~\cite{BHR}, and so improve our discovery reaches slightly.}  Here
and in the following, we use MRS R1 parton distribution
functions~\cite{MRS} and the BASES phase space integration
package~\cite{BASES}.  The cross sections for the different
generations are nearly identical for small $\tan\beta$, since then
$\stauone \approx \stau_R$ and the three sleptons are highly
degenerate.  In the cases of $\selectron_R$ and $\smuon_R$ pair
production, the directly produced slepton may either be stable within
the detector or may decay through virtual Binos via $\slepton_R \to
l_R \tau \stauone$ if kinematically allowed~\cite{AKM}.\footnote{For a
rather narrow range of parameters with $m_{\stauone} < m_{\chi^0_1} <
m_{\slepton_R}$, decays $\slepton_R \to l_R \chi^0_1$ are also
possible~\cite{longLEP}, but we will not consider this scenario here.}
However, in all cases, the signal is a highly ionizing track from a
long-lived charged slepton with no hadronic activity within some
isolation cone.\footnote{If the $\slepton_R$ decays in the detector
with a macroscopic decay length, one might hope to find evidence for
$\slepton_R$ production by searching for leptons with displaced
vertices or for kinks in tracks formed by a $\slepton_R$ and its
$\stauone$ decay product.  Unfortunately, the requirement that the
decay be sufficiently far from the interaction point typically
requires $\slepton_R$--$\stauone$ mass splittings so small that the
leptonic decay products are too soft and the kinks are too small to be
detected.  See, for example, Ref.~\cite{AKM}.}

Highly ionizing tracks may be faked by a real muon that overlaps with
another particle track, which enhances the measured $-dE/dx$.  The
role of the muon may also be played by a hadron that punches through
to the muon chambers.  However, such backgrounds typically arise when
the muon or hadron is part of a jet, and so are drastically reduced by
requiring the highly ionizing particle to be isolated from significant
hadronic activity.  CDF has reported~\cite{CDFtalk} such a search for
highly ionizing tracks. They require the slow particle track to have
$|\eta| \le 0.6$ so that it is detected in the highly shielded central
muon upgrade detector. It is also required to have $-dE/dx$
corresponding to $0.4 < \beta\gamma < 0.85$. The lower limit ensures that
the stau reaches the central muon detector. The upper limit
corresponds roughly to the requirement that the track be at least
twice minimally ionizing.  Such cuts are similar to those imposed in
LEP analyses to eliminate background~\cite{LEP}. They are also
expected to reduce the background at the Tevatron to negligible
levels~\cite{CDFtalk}, and are appropriate for future Tevatron
upgrades~\cite{Stuart}. We will assume that after all of these cuts,
the signal is essentially background-free.\footnote{Note that
measurements of $\beta\gamma = |\vec{p}|/m$ from $-dE/dx$ and $p_T$
from the tracking chambers constrain $m$.  This may be used to further
reduce background, as the signal is peaked at $m=m_{\stauone}$, while
the background is not localized in $m$.}

We may now estimate the number of slepton pair events resulting in
highly ionizing tracks.  In Fig.~\ref{fig:ion_stau}, we plot the cross
section for $\stauone$ pair production for $\sqrt{s}=2\tev$, where we
have required at least one $\stauone$ with $|\eta| \le 0.6$ and
$\beta\gamma < 0.4$, 0.7, 0.85, and $\infty$.  Results for $\sqrt{s} =
1.8\tev$ are similar, and we omit the corresponding figure. We find
that the requirement $\beta\gamma > 0.4$ has little effect on the
signal.  The efficiencies for satisfying the upper bound criterion
$\beta\gamma < 0.85$ are 25\%, 44\%, and 65\% for $m_{\stauone} =
100$, 200, and 300 GeV, respectively. While this degrades the signal
substantially, slow stau cross sections on the order of 0.1 fb are
still possible at slepton masses of 200 GeV.

Finally, we determine the discovery reach for long-lived charged
sleptons.  We sum over three generations.\footnote{In the case where
sleptons $\selectron_R$ and $\smuon_R$ decay via $\slepton_R \to l_R
\tau\stauone$, typically $\Delta m \equiv m_{\slepton_R} -
m_{\stauone} \ll m_{\slepton_R}, m_{\stauone}$, and the $\stauone$ is
produced nearly at rest in the $\slepton_R$ rest frame.  In the lab
frame, the $\stauone$ and $\slepton_R$ velocities are therefore
similar, and so once sleptons $\selectron_R$ or $\smuon_R$ are
produced with low velocity, they result in a highly ionizing track,
irrespective of subsequent decays.}  To crudely simulate the effects
of geometric triggering efficiencies and track quality requirements,
we have included an assumed experimental detection efficiency of 75\%.
With the assumption that the cuts described above reduce the
background to negligible levels, we require, for a given luminosity, a
cross section after including cuts and efficiencies corresponding to 5
or more events for discovery.  The resulting reaches for
representative Tevatron data samples are

\begin{eqnarray}
m_{\slepton_R} = 50\gev &\quad& (\sqrt{s} = 1.8\tev, \ L=110\ipb)
\nonumber \\
m_{\slepton_R} = 110\gev &\quad& (\sqrt{s} = 2\tev, \ L=2\ifb)
\nonumber \\
m_{\slepton_R} = 180\gev &\quad& (\sqrt{s} = 2\tev, \ L=10\ifb)
\nonumber \\
m_{\slepton_R} = 230\gev &\quad& (\sqrt{s} = 2\tev, \ L=30\ifb) \ .
\end{eqnarray}
We see that the discovery reach of 50 GeV for the currently available
Run I data is not competitive with current LEP analyses.  However, for
future Tevatron runs, the discovery reach will be competitive with and
eventually surpass the kinematic limit of LEP II.

\subsection{Gaugino Production}
\label{sec:gaugino}

Aside from right-handed slepton production, the other major SUSY
production mechanisms at the Tevatron are\footnote{Contributions to
gaugino production from $t$-channel squark diagrams are highly
suppressed by large squark masses in the gauge mediation framework and
are omitted in our analysis.}

\begin{equation}
p\bar{p}\to W^{*} \to \chi^\pm_1\chi^0_2 \quad \text{and} \quad
p\bar{p}\to \gamma^*, Z^* \to \chi^\pm_1\chi^\mp_1 \ ,
\label{gaugino}
\end{equation}
where the neutralinos and charginos are Wino-like. The other processes
involving gauginos, $p\bar{p} \to W^* \to \chi^\pm_1\chi^0_1$ and
$p\bar{p}\to Z^* \to \chi^0_{1,2} \chi^0_{1,2}$, are suppressed by
mixing angles in the gaugino limit, and we have checked that,
throughout parameter space, these are not significant relative to
those of Eq.~(\ref{gaugino}).  The cross sections for the processes of
Eq.~(\ref{gaugino}) are given in Fig.~\ref{fig:chi_prod} for both
$\sqrt{s} = 1.8 \tev$ and 2 TeV, where here and in the rest of this
subsection, we neglect neutralino and chargino mixings and take them
to be pure Wino eigenstates.  For equivalent gaugino and slepton
masses, the gaugino cross sections are much larger, as they are
proportional to SU(2) couplings (in contrast to right-handed slepton
production, which is proportional to U(1) couplings in the $\hat{s}
\gg M_Z$ limit) and are not suppressed by the $\beta^3$ behavior of
scalar production.

As in Sec.~\ref{sec:slepton}, we would like to plot cross sections for
events with slow staus.  The decays of gauginos are significantly more
complicated than those of right-handed sleptons.  Of course, the
branching ratios and kinematic distributions of the cascade decays are
completely determined once the fundamental model parameters are fixed.
However, for the present purposes, as we will show, the cascade decays
are determined to a large extent by two rather more meaningful
parameters, namely, the physical Wino and $\slepton_R$ masses.  In
this section, we therefore prefer to highlight this dependence by
adopting a conservative simplifying assumption. (The implications of
these results for the fundamental parameter space will be discussed in
Sec.~\ref{sec:comparison}.)

{}From Fig.~\ref{fig:slM2}, we see that for almost all parameters in
the long-lived stau scenario, and for all parameters when $\nn \ge 3$,
$m_{\slepton_L} < m_{\chi^0_2}, m_{\chi^{\pm}_1}$.  Winos therefore
decay primarily to left-handed sleptons, as the other possible
two-body decays to $W \chi^0_1$, $Z\chi^0_1$, and the ``spoiler mode''
$h^0\chi^0_1$ are all suppressed by mixing angles in the gaugino
region.  The neutral Wino decay modes are then

\begin{eqnarray}
\chi^0_2 &\to& \bar{l}'_L \slepton'_L \to \bar{l}'_L l'_L \chi^0_1 
\to \bar{l}'_L l'_L \bar{l}_R \slepton_R \, , \nonumber \\
\chi^0_2 &\to& \bar{\nu}'_L \snu' \to \bar{\nu}'_L \nu' \chi^0_1 
\to \bar{\nu}'_L \nu' \bar{l}_R \slepton_R \ ,
\label{neutralinodecay}
\end{eqnarray}
and the charged Wino decay modes are

\begin{eqnarray}
\chi^+_1 &\to& \bar{l}'_L \snu' \to \bar{l}'_L \nu' \chi^0_1 
\to \bar{l}'_L \nu' \bar{l}_R \slepton_R \, , \nonumber \\
\chi^+_1 &\to& \nu' \slepton'^*_L \to \nu' \bar{l}'_L \chi^0_1 
\to \nu' \bar{l}'_L \bar{l}_R \slepton_R \ ,
\label{charginodecay}
\end{eqnarray}
where we have omitted states related by charge conjugation. {}From
Figs.~\ref{fig:srM1} and \ref{fig:slM2}, we see also that the
$\slepton'_L$--$\chi^0_2$ and $\slepton_R$--$\chi^0_1$ mass splittings
are never more than $\sim 20\%$ to 30\% and are smaller than this in
much of the parameter space allowed by perturbativity.  In the limit
that these states are highly degenerate, there are only two energetic
final state particles in the Wino rest frame for any of the possible
decays: the lepton from $\slepton'_L$ or $\snu'_L$ decay, and the
final state $\slepton_R$.  We therefore make the following
simplification: in the rest frame of the parent gaugino, we assume
that the only energetic particles are one lepton and the final
$\slepton_R$. The $\slepton_R$ velocity in this rest frame is then
fixed by

\begin{eqnarray}
\beta\gamma|_{\text{rest}} = \frac{M_2^2 - m_{\slepton_R}^2}{2 M_2
m_{\slepton_R}} \ ,
\end{eqnarray} 
where $\beta$ is the slow slepton's velocity and $\gamma =
(1-\beta^2)^{-1/2}$. The cross section for slow $\slepton_R$
production is then completely determined by $M_2$ and
$m_{\slepton_R}$.

In Figs.~\ref{fig:ion_wino5}--\ref{fig:ion_wino3}, we show slow
$\slepton_R$ production cross sections for three values of
$m_{\slepton_R}/M_2$ and $\sqrt{s}=2\tev$, where we sum over the three
slepton generations and employ the simplifying assumption described
above.  (Again, results for $\sqrt{s} = 1.8\tev$ are fairly similar,
and the corresponding figures are not presented.)  To account for the
roughly $2\pi$ coverage of the central muon upgrade detector, we have
assumed an $\eta$ cut efficiency of 50\%. {}From these figures, we see
that for fixed gaugino mass, the cross sections are maximized for
large $m_{\slepton_R}$, that is, small
$\beta\gamma|_{\text{rest}}$. For $M_2 = 400 \gev$, for example, the
$\beta\gamma < 0.85$ efficiency is 62\%, 32\%, and 7\% for
$m_{\slepton_R}/M_2 = 0.5$, 0.4, and 0.3, respectively.  This can be
understood as a consequence of the low velocity of the heavy Winos in
the lab frame, which implies that the $\slepton_R$ velocity in the lab
frame is mostly determined by its velocity in the Wino rest frame.
Note that our simplifying assumption implies that we have taken the
maximal possible value of $\beta\gamma|_{\text{rest}}$.  In regions of
the parameter space where $\slepton_L$ and $\chi^0_2$ (and
$m_{\slepton_R}$ and $m_{\chi^0_1}$) are non-degenerate at the 20\% to
30\% level, additional leptons can have significant energies.  This
reduces $\beta\gamma|_{\text{rest}}$, and therefore increases the slow
particle cross sections.  Our approximation is therefore conservative.

To determine the discovery reach for long-lived sleptons from gaugino
production, we include a 75\% experimental efficiency and require 5 or
more events for discovery, as in Sec.~\ref{sec:slepton}.  The
resulting discovery reaches in $M_2$ for our representative set of
luminosities are
 
\begin{eqnarray}
M_2 = \text{120 to 150 GeV} &\quad& 
(\sqrt{s} = 1.8\tev, \ L=110\ipb)
\nonumber \\
M_2 = \text{220 to 280 GeV} &\quad& 
(\sqrt{s} = 2\tev, \ L=2\ifb)
\nonumber \\
M_2 = \text{270 to 340 GeV} &\quad& 
(\sqrt{s} = 2\tev, \ L=10\ifb)
\nonumber \\
M_2 = \text{310 to 390 GeV} &\quad& 
(\sqrt{s} = 2\tev, \ L=30\ifb) \ ,
\end{eqnarray}
where the ranges are for $m_{\slepton_R}/M_2 = 0.3$ to 0.5. Because
the signal is background-free, the discovery reaches grow rapidly with
luminosity, and future Tevatron runs substantially extend our reach in
SUSY parameter space.

\section{Dimuon-like Events} 
\label{sec:dimuon}

As discussed in the previous section, a long-lived charged slepton
appears in detectors as a muon.  Thus, even if the velocity of such a
slepton is too high to be seen as a highly ionizing track, slepton
production can lead to an excess of events with muons and an apparent
violation of lepton universality.  In this section, we discuss the
simplest example of such an excess, namely, an apparent excess in the
number of dimuon events from right-handed slepton pair production.

The Drell-Yan production of long-lived $\stauone$ pairs will appear as
an event with essentially no jet activity and two high energy
muon-like tracks with no $\mpt$. The production of right-handed
sleptons of the first two generations may also lead to this signature.
For $\selectron_R$ and $\smuon_R$, there are three possibilities: (1)
they may be stable within the detector; (2) they may decay via
$\slepton_R \to l_R \tau \stauone$ to leptons that are so soft that
they cannot be clearly identified in the detector; or (3) they may
decay to the NLSP and detectable leptons.  In the first two cases,
which are applicable when $\tan\beta$ is small and $\selectron_R$,
$\smuon_R$ and $\stauone$ are nearly degenerate, the pair production
of smuons and selectrons contributes to the dimuon event sample.  We
will assume this to be the case in this section.  This assumption is
conservative, in the sense that in the third scenario, $\selectron_R$
and $\smuon_R$ pair production contributes to spectacular multi-lepton
signals, which are virtually background-free.  Such signals will be
discussed in the following section.

The backgrounds to dimuon signals with little $\mpt$ have been studied
in the context of searches at CDF for new neutral gauge bosons via
their decays $Z'\to \mu^+\mu^-$~\cite{Zprime}.  The dominant
background is from genuine muon pair production $p\bar{p}
\to\mu^+\mu^-$.\footnote{Of the remaining backgrounds, the largest is
cosmic rays~\cite{Zprime}.  This cross section is estimated to be less
than $1\fb$ for invariant dimuon masses above 200 GeV~\cite{Zprime},
and is negligible relative to the muon pair background if
well-understood.}  Before cuts, the cross section for this process is
several orders of magnitude larger than that for slepton pair
production. However, the signal to background ratio may be improved by
requiring both muon-like tracks to have $p_T$ greater than some fixed
$\ptcut$.  In Fig.~\ref{fig:dimuon}, we plot the cross sections for
right-handed slepton pair and muon pair production at the Tevatron
with $\sqrt{s}=2$~TeV as a function of $\ptcut$. Following
Ref.~\cite{Zprime}, we require that at least one muon satisfy $|\eta|
\le 0.6$, and the other have $|\eta| \le 1.0$ so that its $p_T$ may be
well-measured in the central tracking chamber.  As is evident in
Fig.~\ref{fig:dimuon}, the $p_T$ cut is highly effective for $\ptcut
\agt 100\gev$.

To estimate the discovery potential for a dimuon excess, we look for
an apparent violation of lepton universality in the ratio
$\sigma(\mu^+ \mu^-) / \sigma(e^+ e^-)$.  By considering this ratio
instead of the absolute $\mu^+ \mu^-$ cross section, large systematic
uncertainties from luminosity measurements and renormalization scale
dependence in next-to-leading order calculations are removed.  The
statistical significance of deviations in this ratio, in units of
standard deviation $\sigma$, is given by $S/\sqrt{2B}$, where $S$ is
the number of slepton pair events, $B$ is the number of muon pair
events, and we have neglected differences in $e$ and $\mu$ acceptances
and systematic errors in determining these acceptances.  For a given
slepton mass, we calculate the ratio $S/\sqrt{2B}$ as a function of
$\ptcut$, and optimize the choice of $\ptcut$ to maximize
$S/\sqrt{2B}$. We assume an overall muon trigger efficiency of 75\%,
and sum over the three slepton generations for $S$.  In
Table~\ref{table:dimuon}, we show the optimized value of $S/\sqrt{2B}$
and the optimal choice of $\ptcut$ for various values of
$m_{\stauone}$ and $L$.  The optimal $\ptcut$ is typically $\ptcut
\sim m_{\stauone}$.

While many systematic errors are eliminated by considering the
ratio $\sigma(\mu^+ \mu^-) / \sigma(e^+ e^-)$, systematic
uncertainties in muon and electron triggering and acceptance
efficiencies, of course, are not.  The relative efficiencies may be
determined, for example, by isolating leptons from $Z$ decays and
assuming lepton universality.  However, if such efficiencies are
uncertain at or near the level of $S/B$, they will seriously degrade
one's ability to identify an excess of muons.  We have therefore
presented values of $S/B$ for each $m_{\stauone}$ in
Table~\ref{table:dimuon}; typical values are 10\% to 20\%.  Relative
efficiencies must be known to significantly better than these values
for a muon excess to be detected. In the following, we will assume
this to be the case and that the dominant uncertainty in detecting a
muon excess is statistical.

\begin{table}[t]
\begin{center}
\begin{tabular}{cccccr}
{$m_{\stauone}$} & {$2\ifb$} & {$10\ifb$} & {$30\ifb$} & 
{Optimal $\ptcut$} & {$S/B$}
\\ \hline
{80~GeV}  & {1.6$\sigma$} & {3.7$\sigma$} & {6.4$\sigma$} & {100~GeV}
& {22\%} \\
{100~GeV} & {1.0$\sigma$} & {2.3$\sigma$} & {4.0$\sigma$} & {120~GeV}
& {19\%} \\
{120~GeV} & {0.7$\sigma$} & {1.4$\sigma$} & {2.6$\sigma$} & {140~GeV}
& {16\%} \\
{140~GeV} & {0.4$\sigma$} & {1.0$\sigma$} & {1.7$\sigma$} & {160~GeV}
& {13\%} \\
{160~GeV} & {0.3$\sigma$} & {0.6$\sigma$} & {1.1$\sigma$} & {180~GeV}
& {11\%} \\
{180~GeV} & {0.2$\sigma$} & {0.4$\sigma$} & {0.7$\sigma$} & {200~GeV}
& {9\%} \\ 
\end{tabular}
\caption{Statistical significances of deviations in the ratio
$\sigma(\mu^+ \mu^-) / \sigma(e^+ e^-)$ from contributions of
long-lived charged slepton pair production to dimuon events for
integrated luminosities $L=2$, 10, and $30 \ifb$ and $\sqrt{s} =
2\tev$ at the Tevatron.  The signal from slepton pair production is
summed over the three generations. The cut parameter $\ptcut$ is
chosen to maximize the significance; the optimal values used are
shown.  Values of $S/B$ after cuts are as shown.}
\label{table:dimuon} 
\end{center} 
\end{table}

The discovery reach for a $3\sigma$ excess in dimuon events is below
50 GeV for Run I, and for future Tevatron upgrades is
  
\begin{eqnarray}
m_{\slepton_R} =70\gev &\quad& (\sqrt{s} = 2\tev, \ L=2\ifb)
\nonumber \\
m_{\slepton_R} = 90\gev &\quad& (\sqrt{s} = 2\tev, \ L=10\ifb)
\nonumber \\
m_{\slepton_R} = 110\gev &\quad& (\sqrt{s} = 2\tev, \ L=30\ifb) \ .
\end{eqnarray}
We see that this discovery mode is clearly less powerful than the
previous signal discussed in Sec.~\ref{sec:ionization}, and we will
see that it is also weaker than that of the following section.  In
fact, for an integrated luminosity of $L = 2\ifb$, it is not possible
to achieve a 3$\sigma$ excess, given current bounds from LEP.
However, this mode may still be useful as a supplementary signal.  For
example, if highly ionizing tracks consistent with a certain charged
particle mass are seen, the discovery of even a small but
quantitatively consistent dimuon excess could be an interesting
confirmation of the supersymmetric interpretation of these exotic
signatures.  Note also that it may be possible to improve the
sensitivity of the dimuon channel by loosening the $\eta$ cuts,
otherwise improving acceptance, or by combining data from the two
detectors.

Given the rather weak results for the Tevatron, we consider the
possibility of observing a dimuon excess from $pp\to \slepton_R
\slepton_R^*$ at the LHC with $\sqrt{s}=14\tev$.  We require $\eta \le
2.2$ for both sleptons, corresponding approximately to the muon
coverage of both LHC experiments~\cite{LHC}, and, as above, assume a
75\% experimental detection efficiency and sum over 3 generations.  In
Table~\ref{table:dimuonLHC} we present results for integrated
luminosities of $10\ifb$ (one year, low luminosity), $100\ifb$ (one
year, high luminosity), and $600\ifb$ (corresponding to multi-year
runs and/or combined data sets from the two detectors).  Not
surprisingly, the results are much improved.  For example, for
$L=100\ifb$, a $3\sigma$ excess will be discovered for $m_{\slepton_R}
\alt 300\gev$, where we have again assumed that systematic
uncertainties in acceptance are small compared to statistical
uncertainties.  Of course, at the LHC, a number of other signals of
gauge-mediated supersymmetry are likely to be evident, and even squark
and gluino cross sections may be substantial.

\begin{table}[t]
\begin{center}
\begin{tabular}{cccccr}
{$m_{\stauone}$} & {$10\ifb$} & {$100\ifb$} & {$600\ifb$} & 
{Optimal $\ptcut$} & {$S/B$}
\\ \hline
{100~GeV} & {$8.1\sigma$} & {$26\sigma$} & {$63\sigma$} & {150~GeV}
& 28\% \\
{200~GeV} & {$2.4\sigma$} & {$7.7\sigma$} & {$19\sigma$} & {250~GeV}
& 20\% \\
{300~GeV} & {$1.1\sigma$} & {$3.4\sigma$} & {$8.3\sigma$} & {350~GeV}
& 16\% \\
{400~GeV} & {$0.5\sigma$} & {$1.7\sigma$} & {$4.1\sigma$} & {450~GeV}
& 13\% \\
{500~GeV} & {$0.3\sigma$} & {$0.9\sigma$} & {$2.3\sigma$} & {550~GeV}
& 11\% \\
{600~GeV} & {$0.2\sigma$} & {$0.5\sigma$} & {$1.3\sigma$} & {650~GeV}
& 9\% \\
\end{tabular}
\caption{Same as in Table~\protect\ref{table:dimuon}, but for
integrated luminosities $L=10$, 100, and $600 \ifb$ and $\sqrt{s} =
14\tev$ at the LHC. }
\label{table:dimuonLHC}
\end{center}
\end{table}

\section{Multi-Lepton Signals}
\label{sec:multi-lepton}

If stau NLSPs are misidentified as muons, supersymmetric particle
production can lead to events that appear to have large numbers of
leptons in the final state.  Here we focus on events with 5 or more
isolated leptons and very little hadronic activity.  Assuming that the
probability for jets and photons to fake leptons is well under
control, the backgrounds to such events are small, and so even a few
signal events may constitute a discovery signal. Such multi-lepton
states may result from either right-handed slepton or gaugino
production, and, as in Sec.~\ref{sec:ionization}, we consider these in
turn.

\subsection{Slepton Production}
\label{sec:2ndslepton}

For small $\tan\beta$, right-handed slepton pair production leads to
dimuon events, as analyzed in Sec.~\ref{sec:dimuon}.  However, for
large $\tan\beta$, the intergenerational splitting from the tau Yukawa
coupling may be substantial, and selectrons and smuons may decay to
NLSPs and sufficiently energetic leptons to be detected as
multi-lepton events.

To understand what values of $\tan\beta$ are required for this
scenario, let us consider the various contributions to the
intergenerational mass difference. Left-right stau mixing gives an
important contribution,

\begin{eqnarray}
\Delta m_{\text{LR}} \simeq \frac{m_\tau^2\mu^2\tan^2\beta}
{2m_{\stau_R}(m_{\stau_L}^2-m_{\stau_R}^2)} \ .
\end{eqnarray}
In addition, the splitting receives contributions from RG evolution.
Roughly speaking, this effect is proportional to $\ln(M/ \mmssm)$ and
the beta function for $M_{\stau_R}^2$, and hence may be estimated as

\begin{eqnarray}
\Delta m_{\rm RG} &\sim& \frac{1}{2m_{\stau_R}}\times
\frac{1}{4\pi^2} y_\tau^2 \times {\cal O} \left[ m_{\stau_L}^2 
\ln(M/ \mmssm) \right] \nonumber \\ 
&\sim& \frac{1}{2\pi^2} \times {\cal O} \left[\frac{m_{\stau_L}^4}
{v^2 \mu^2} \ln(M/ \mmssm)\right] \times \Delta m_{\text{LR}} \ ,
\end{eqnarray}
where $y_\tau$ is the $\tau$ Yukawa coupling constant, $v=246\gev$,
and in the last line we have assumed large $\tan\beta$. $\Delta m_{\rm
RG}$ may be sizeable compared to $\Delta m_{\text{LR}}$ when
$m_{\stau_L}$ is large relative to the electroweak scale. In any case,
both contributions reduce $m_{\stauone}$ and they are both enhanced
when $\tan\beta$ is large. As a result, mass differences between
$\stauone$ and the other right-handed sleptons can become large for
high $\tan\beta$.  In Fig.~\ref{fig:massdiff}, we plot the mass
splitting $\Delta m \equiv m_{\selectron_R} - m_{\stauone} \simeq
m_{\smuon_R} - m_{\stauone}$ as a function of $m_{\stauone}$ for fixed
$\nn=3$, $\mmess=10^5 \gev$, $\mu > 0$, and various $\tan\beta$.  If
$\tan\beta$ is small, $\Delta m < m_{\tau}$, and all three
right-handed sleptons are stable in the detector. However, for
$\tan\beta = 30$, for example, mass splittings of 40 GeV or larger are
possible, and $\selectron_R$ and $\smuon_R$ may decay inside the
detector to electrons and muons that are likely to be observable.

We now consider the scenario in which right-handed selectron and smuon
pair production leads to final states $(ee,\mu\mu)
\tau\tau\stauone\stauone$.  For high $\tan\beta$ and correspondingly
large mass splittings $\Delta m$, the resulting electrons and muons,
and even the leptonic decay products of the tau leptons, are typically
energetic enough to be detected. Demanding that at least one $\tau$
decay leptonically, this leads to signals of either 5 energetic,
isolated leptons and the hadronic decay products of a $\tau$, or 6
energetic, isolated leptons with no significant hadronic activity. We
expect the backgrounds to such signals to be extremely small. On the
other hand, for more moderate values of $\tan\beta$, the leptons are
typically softer, and the number of events with 5 or 6 observable
leptons is reduced. While it may be possible to lower the requirement
on lepton energy, significant backgrounds may then enter.

To obtain an estimate of the possible reach of this mode, we assume,
for simplicity, that $\tan\beta$ is sufficiently large, for example,
$\tan\beta \agt 30$, such that to a good approximation, all leptonic
decay products are energetic enough to be detected.  We take the
signal to be 5 or more isolated leptons with no significant hadronic
activity, other than that consistent with the hadronic decay products
of a $\tau$.  Slepton pair production then contributes to the signal
when at least one $\tau$ decays leptonically.  With these assumptions,
we expect the signal to be essentially background-free and we
therefore require a discovery signal of 5 events.  To roughly
incorporate the effects of detector acceptances, we include an overall
experimental efficiency of 50\%.  {}From Fig.~\ref{fig:sl_prod}, but
for 2 generations, we find the following discovery reaches for slepton
masses:

\begin{eqnarray}
m_{\slepton_R} = 60\gev &\quad& (\sqrt{s} = 1.8\tev, \ L=110\ipb)
\nonumber \\
m_{\slepton_R} = 130\gev &\quad& (\sqrt{s} = 2\tev, \ L=2\ifb)
\nonumber \\
m_{\slepton_R} = 170\gev &\quad& (\sqrt{s} = 2\tev, \ L=10\ifb)
\nonumber \\
m_{\slepton_R} = 210\gev &\quad& (\sqrt{s} = 2\tev, \ L=30\ifb) \ .
\end{eqnarray}

As noted above, these estimates of the discovery reach are valid only
for large $\tan\beta$, and are degraded for lower $\tan\beta$.  A
quantitative determination of the dependence of the reach on
$\tan\beta$, requiring a careful study of backgrounds and the
optimization of cuts, merits further study, but is beyond the scope of
the present analysis.

\subsection{Gaugino Production}
\label{sec:2ndgaugino}

As discussed in Sec.~\ref{sec:gaugino}, in the stau NLSP scenario,
Wino-like charginos and neutralinos typically decay through
left-handed sleptons, with cascade decays given in
Eqs.~(\ref{neutralinodecay}) and (\ref{charginodecay}).  Assuming that
right-handed sleptons appear in the detector as muons, the final state
therefore contains many charged leptons and no jets.  For scenarios
with gauginos that are highly degenerate with sleptons, some of the
real leptons in the final state may be too soft to detect.  However,
this applies only to points extremely close to the $m_{\stauone} =
M_1$ boundary of Fig.~\ref{fig:MmessN5}, and we will assume that this
is not the case. Gaugino production then results in the spectacular
signal of from 5 to 7 charged leptons.\footnote{We assume here that
$\selectron_R$ and $\smuon_R$ do not further decay through $\slepton_R
\to l_R\tau\stauone$.  If such decays result in detectable leptons,
the signals are even more spectacular.}

As in Sec.~\ref{sec:2ndslepton}, we take the signal to be 5 or more
isolated leptons with little hadronic activity.  We demand 5 events
for a discovery signal, and again include an overall experimental
detector efficiency of 50\%.  For the 7 lepton events, this is likely
to be a rather conservative estimate, as only 5 or more need be
detected. We may then determine the discovery reach for multi-lepton
signals in this scenario from Fig.~\ref{fig:chi_prod}.  The discovery
reaches for our representative data samples are

\begin{eqnarray}
M_2 = 190\gev &\quad& (\sqrt{s} = 1.8\tev, \ L=110\ipb)
\nonumber \\
M_2 = 310\gev &\quad& (\sqrt{s} = 2\tev, \ L=2\ifb)
\nonumber \\
M_2 = 370\gev &\quad& (\sqrt{s} = 2\tev, \ L=10\ifb)
\nonumber \\
M_2 = 410\gev &\quad& (\sqrt{s} = 2\tev, \ L=30\ifb) \ .
\end{eqnarray}
The multi-lepton signal from Wino production is thus one of the most
sensitive probes, and probes chargino and neutralino masses far into
the ranges typically expected for Wino masses.

\section{Comparison of Results}
\label{sec:comparison}

In the previous 3 sections, we have considered a variety of SUSY
discovery signals in the long-lived stau NLSP scenario.  They are:

\begin{itemize}
\item[(A)] Highly ionizing tracks from slepton pair 
production (Sec.~\ref{sec:slepton}).
\item[(B)] Highly ionizing tracks from Wino pair 
production (Sec.~\ref{sec:gaugino}).
\item[(C)] An excess of dimuon events from slepton pair 
production (Sec.~\ref{sec:dimuon}).
\item[(D)] Multi-lepton signals from slepton pair production 
(Sec.~\ref{sec:2ndslepton}).
\item[(E)] Multi-lepton signals from Wino pair production 
(Sec.~\ref{sec:2ndgaugino}).
\end{itemize}

For each of these signals, we have obtained discovery reaches in terms
of physical right-handed slepton and/or Wino masses by the analyses
described above.  We have found that the discovery reach for slepton
masses from (A), (C), and (D) are numerically weaker than those for
gaugino masses from (B) and (E). However, in typical gauge mediation
scenarios, the Winos are predicted to be substantially heavier than
the right-handed sleptons.  It is therefore of interest to determine
what the relative strengths of these probes are in terms of the
fundamental parameter space of the gauge mediation model.

In Figs.~\ref{fig:gm3_run1}--\ref{fig:gm4_tot}, we summarize all of
our previous results by presenting discovery reaches in the
fundamental parameter space for various integrated luminosities.  We
display these reaches in the $(\mmess, \lmess)$ plane, and fix the
remaining three parameters $\tan\beta = 3$, $\mu>0$, and $\nn$.  The
number of messenger representations is $\nn=3$ for
Figs.~\ref{fig:gm3_run1} and \ref{fig:gm3_tot}; in
Figs.~\ref{fig:gm4_run1} and \ref{fig:gm4_tot}, we illustrate the
$\nn$ dependence of these results by taking $\nn=4$. For each point in
the plane, the complete superpartner spectrum is specified, and, in
these figures, $D$-terms for scalar masses and all mixing effects are
included in calculating physical masses and production cross sections.
In each figure, we shade the area in which $m_{\stauone} >
m_{\chi^0_1}$; in the remaining region, the stau is the NLSP.

To guide the eye in relating the fundamental model parameters to more
physical parameters, we plot dotted contours of constant
$m_{\stauone}$.  Also, since $M_i/g_i^2$ is invariant under one-loop
RG evolution and the weak scale coupling constants $g_i(\mmssm)$ are
known, the weak scale gaugino masses $M_i(\mmssm)$ are proportional to
$\lmess$, independent of $\mmess$.  In particular, the SU(2)$_L$
gaugino mass is given by

\begin{equation}  
M_2(\mmssm)=\nn \, \lmess \, \frac{g_2^2(\mmssm)}{16\pi^2} \ .
\label{m2lambda}
\end{equation}
We therefore also give vertical axis labels for $M_2(\mmssm)$ with
$\mmssm=1\tev$.  This axis gives approximately the physical masses of
the Wino-like chargino and neutralino.
 
For each point in parameter space with a stau NLSP, we determine the
potential for SUSY discovery for each signal according to the analyses
described in Secs.~\ref{sec:ionization}--\ref{sec:multi-lepton}.
Contours for signals (A), (C), and (D) therefore roughly follow
contours of constant right-handed slepton mass, and contours for
signal (E) roughly follow contours of constant chargino and neutralino
mass.  For signal (B), the contours also roughly follow chargino and
neutralino mass contours, but rise slightly relative to these contours
as $\mmess$ increases: in this direction, $m_{\slepton_R}$ increases,
and the signal is strengthened, as discussed in
Sec.~\ref{sec:gaugino}.

We begin by presenting results in Figs.~\ref{fig:gm3_run1} and
\ref{fig:gm4_run1} for the present Tevatron data sample of
$L=110\ipb$ at $\sqrt{s} = 1.8\tev$.  Several comments are in order.
First, we have assumed $\tan\beta=3$, and so slepton production does
not result in multi-lepton events.  Signal (D) is therefore not
applicable.  Second, signals (A) and (C) are too weak to appear in the
plots.  These signals are both from slepton production, and are
essentially dependent on only the right-handed slepton mass.  Given
only the current Run I data, they are therefore weaker than the
current LEP bound of $m_{\stau_R} > 75\gev$, independent of model
assumptions. For signal (B), the reaches plotted are conservative, as
discussed in Sec.~\ref{sec:gaugino}.  Finally, signal (E),
multi-lepton events from Wino production, requires that the leptons
from $\tilde{W} \to \slepton_L$ and $\tilde{B} \to \slepton_R$
transitions be energetic enough to be detected.  For $\nn=3$ and 4,
this is always true for the former decay in the stau NLSP region.
However, the latter decay may result in soft leptons, and the bound
from signal (E) is therefore weakened for points very close to the
boundary of the shaded region.  We have not included this effect in
the figures.
 
For the particular parameters we have chosen, a number of interesting
conclusions may be drawn.  Given the current Run I data, the slepton
production signals (A) and (C) are superseded by the requirement that
the parameters yield the stau NLSP scenario in the first place.  Thus,
for low $\tan\beta$, only the gaugino production signals (B) and (E)
provide non-trivial probes of the parameter space. However, even given
only the current Tevatron data sample, signals (B) and (E) do probe
regions of parameter space not excluded by the current LEP bound
$m_{\stau_R} > 75\gev$~\cite{LEPcombined}.  For $\nn=3$, we see from
Fig.~\ref{fig:gm3_run1} that signal (B) is sensitive to parameter
regions beyond this bound, and signal (E) probes regions with
$m_{\stauone} \approx 110\gev$, which are even beyond the ultimate
kinematic reach of LEP II.  For $\nn=4$, the ratio $m_{\stauone}/M_2$
is reduced, and the strength of gaugino production mechanisms relative
to slepton mechanisms is slightly weakened.  However, as seen in
Fig.~\ref{fig:gm4_run1}, signal (B) is still competitive with the
current LEP bound, and signal (E) will significantly extend the reach
of LEP II for certain values of $\mmess$.

In the context of the stau NLSP scenario, if no signals are found in
analyses of the available Tevatron data, current bounds on Wino masses
will be extended to roughly 200 GeV.  In addition, in the framework of
the particular model we have described, an indirect bound on
$m_{\stauone}$ may be set.  For example, for $\nn=3$, we see from
Fig.~\ref{fig:gm3_run1} that stau masses below $\approx 90\gev$ would
be excluded by the absence of signal (E).  Of course, such indirect
bounds on stau masses rely heavily on the specific model assumptions
described in Sec.~\ref{sec:model}.

In Figs.~\ref{fig:gm3_tot} and \ref{fig:gm4_tot}, we present results
for the same model parameters, but for future Tevatron runs with
$\sqrt{s} = 2\tev$ and $L = 2$, 10, and $30\ifb$.  Again, as we have
assumed $\tan\beta = 3$, discovery reaches from (D) multi-lepton
signatures of slepton pair production are inapplicable.  Signal (C) is
again weak relative to the others. For integrated luminosities of
$2\ifb$ or more, direct probes of slepton masses from signal (A) and
indirect probes from signals (B) and (E) all exceed $m_{\stauone} \sim
100\gev$, and SUSY discovery in these modes is possible, even given
potential LEP II bounds on the stau mass.  For $L=2\ifb$, the
multi-lepton gaugino production signal (E) is dominant, and indirectly
excludes stau masses below $\sim 120\gev$.  However, for increasing
luminosity, the relative importance of the slow slepton signal (A)
increases.  For $L=10\ifb$ and $L=30\ifb$, signal (A) becomes the most
sensitive probe, with sensitivity to regions of parameter space with
Wino masses in excess of 400 to 600 GeV.

\section{Conclusions}
\label{sec:conclusions}

Gauge-mediated supersymmetry breaking models provide an elegant
solution to one of the most puzzling issues in supersymmetry, namely,
the supersymmetric flavor problem.  The phenomenology of these models
is to a large extent dominated by the character of the NLSP.  In the
most popular gauge mediation scenarios, there are four options, as the
NLSP may be either the neutralino or stau, and may be either stable or
unstable in detectors.  Three of these scenarios have received fairly
extensive attention in the past. In this study, we have given for the
first time a detailed treatment of prospects for SUSY discovery at the
Tevatron in the fourth scenario, in which the NLSP is a long-lived
stau. As discussed in Sec.~\ref{sec:model}, this scenario is realized
in a wide range of parameter space in models with messenger sectors
with $\nn \ge 2$.

In such scenarios, all superparticles decay ultimately to the lighter
stau $\stauone$, which appears in collider detectors as a (possibly
slow) muon.  We have analyzed the prospects for discovering SUSY by
detecting highly ionizing tracks from slow $\stauone$ production, an
apparent excess in dimuon events from stau pairs, and the appearance
of spectacular multi-lepton events from superpartner pair production
followed by decays.

Of the signals considered, the dimuon excess is least promising, as it
suffers from the large background of genuine $\mu^+\mu^-$ production.
Its use appears to be limited to confirming the supersymmetric
interpretation of other, more sensitive signals.  Note, however, that
a dimuon excess is only the simplest example of an apparent violation
of lepton universality from long-lived $\stauone$ production.  Many
other possibilities exist.  If a statistically significant excess of
$\mu$ over $e$ events appears in any event sample, we strongly
encourage the investigation of the long-lived charged slepton
hypothesis, either by looking for highly ionizing slow tracks in this
event sample, or by forming distributions of kinematic variables under
the assumption that the muons are in fact sleptons of some given mass.

The remaining signals are sufficiently spectacular that they are
essentially background-free after the cuts we have detailed in
Secs.~\ref{sec:ionization} and \ref{sec:multi-lepton}.  We have
analyzed various data samples, beginning with Run I data with
$\sqrt{s} = 1.8\tev$ and integrated luminosity $L=110\ipb$.  With this
current data, we find that the sensitivity from slepton production is
not competitive with the current bound on long-lived staus of
$m_{\stau_R} \agt 75\gev$ from combined LEP analyses.  However, Winos
with masses beyond the LEP kinematic limit may be produced in
significant numbers at the Tevatron.  Wino pair production therefore
probes regions of parameter space beyond the current LEP bound, and
even beyond the possible ultimate kinematic reach of LEP II.  The
analysis of currently available Tevatron data therefore provides
interesting new probes of the gauge mediation parameter space.

Given the great interest in future Tevatron upgrades, we then
considered what improvements may be expected from future Tevatron
data.  As the backgrounds to the signals we consider are extremely
suppressed, we find that significant improvements may be expected.
For future Tevatron runs with integrated luminosities of 2, 10, and
$30\ifb$, we estimate that SUSY may be discovered for right-handed
slepton masses of 110, 180, and 230 GeV and Wino masses of 310, 370,
and 420 GeV, respectively.  These new and spectacular signals are
therefore sensitive to much of the typically expected superpartner
mass range and provide a promising avenue for SUSY searches at
Tevatron runs in the near future.

\acknowledgements

We are especially grateful to D.~Stuart for many helpful conversations
concerning experimental issues in detecting highly ionizing tracks. We
also thank K.~Agashe, B.~Dutta, H.~Frisch, M.~Graesser, T.~Han, and
I.~Hinchliffe for discussions, and the Aspen Center for Physics for
hospitality during the inception of this work. This work was supported
in part by the Director, Office of Energy Research, Office of High
Energy and Nuclear Physics, Division of High Energy Physics of the
U.S. Department of Energy under Contract DE--AC03--76SF00098, and in
part by the National Science Foundation under grant PHY--95--14797.

\newpage

\begin{figure}[t]
\centerline{\epsfxsize=0.554\textwidth \epsfbox{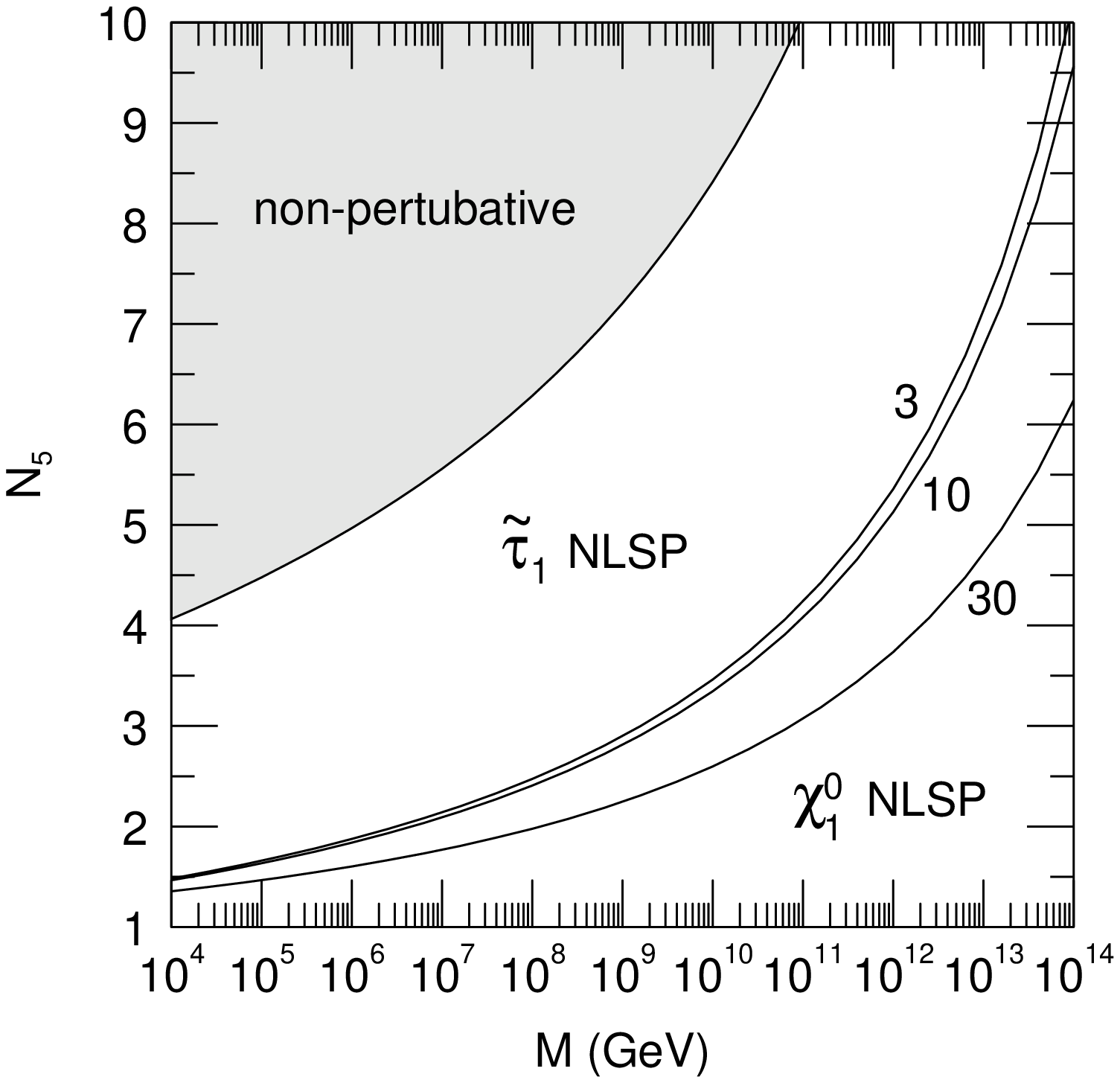}}
\vspace*{0.222in}
\caption{Contours of $M_{\stau_R}(\mmssm) = M_1(\mmssm)$
for $\mmssm=1\tev$ and $\tan\beta = 3$, 10, and 30.  In the region
above the contours, $\stauone$ is the NLSP; in the region below,
$\chi^0_1$ is the NLSP.  In the shaded region, gauge coupling
constants become non-perturbative below the GUT scale under two-loop
RG evolution. }
\label{fig:MmessN5}
\end{figure}

\begin{figure}[t]
\centerline{\epsfxsize=0.554\textwidth \epsfbox{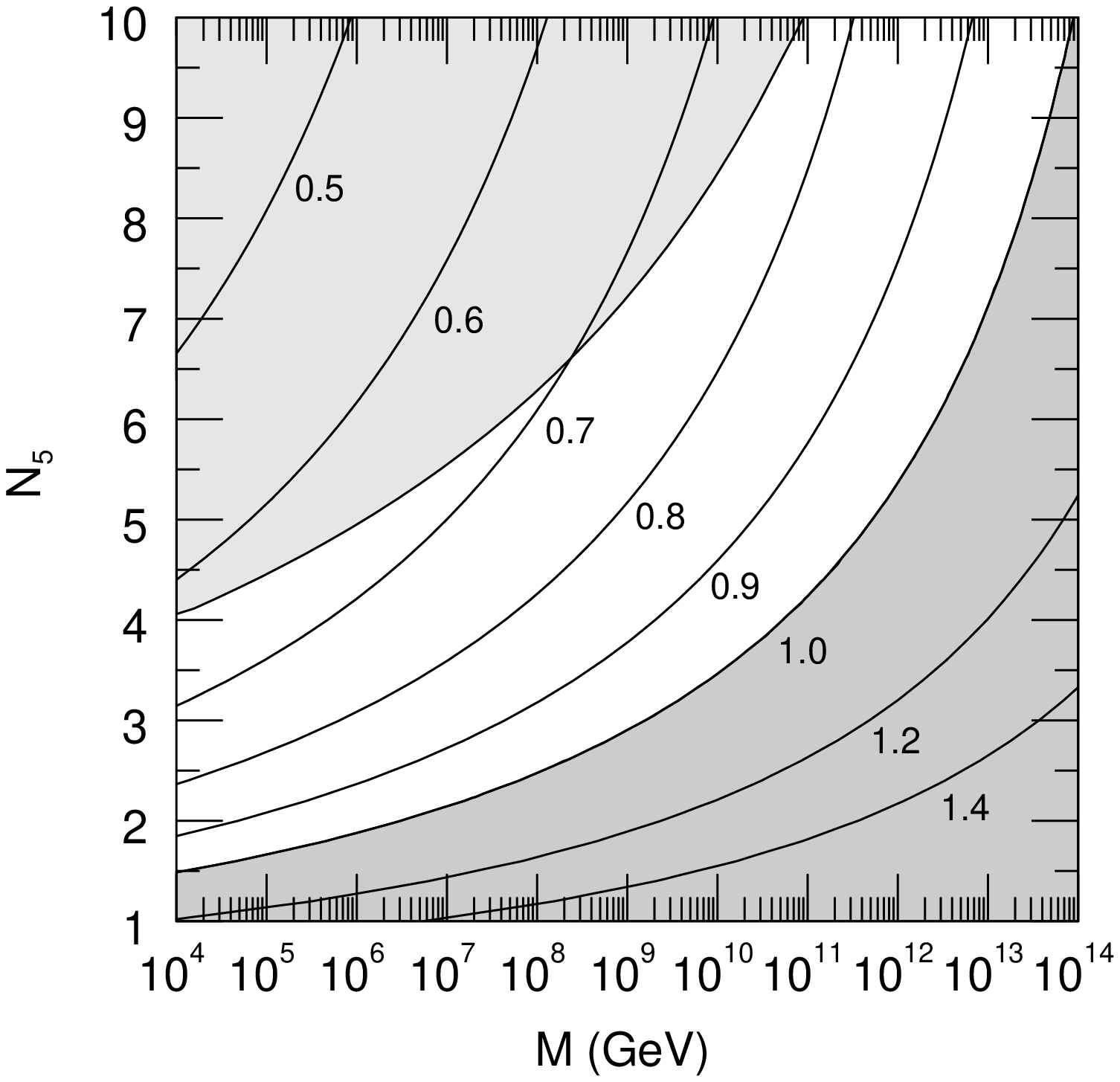}}
\vspace*{0.222in}
\caption{Contours of the ratio of soft SUSY breaking parameters 
$M_{\stau_R}(\mmssm)/M_1(\mmssm)$ for $\tan\beta=3$ and
$\mmssm=1\tev$. In the lower shaded region, $M_{\stau_R}(\mmssm) >
M_1(\mmssm)$. In the upper shaded region, gauge coupling constants
become non-perturbative below the GUT scale. }
\label{fig:srM1}
\end{figure}

\begin{figure}[t]
\centerline{\epsfxsize=0.554\textwidth \epsfbox{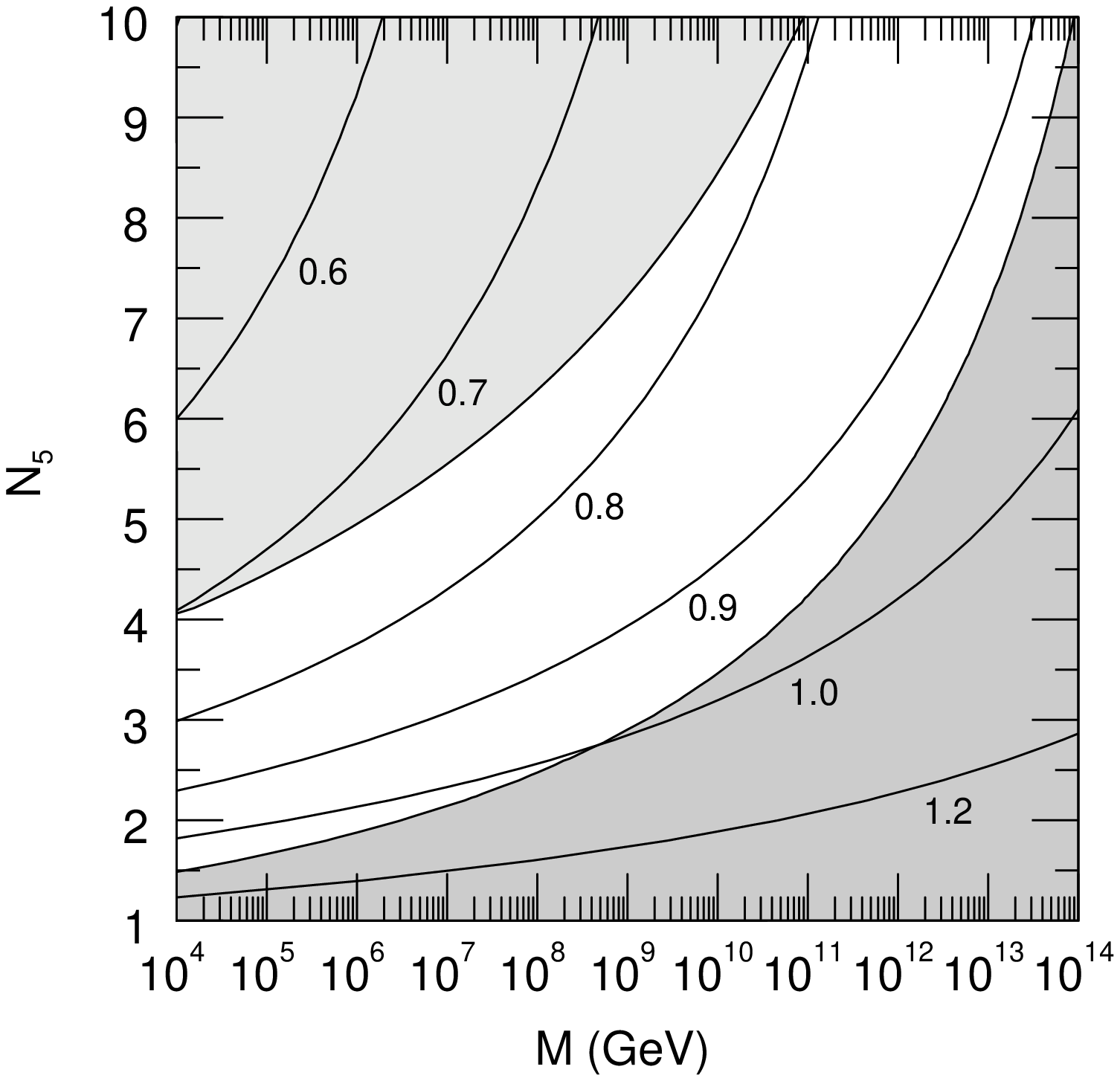}}
\vspace*{0.222in}
\caption{Contours of the ratio of soft SUSY breaking parameters 
$M_{\slepton_L}(\mmssm)/M_2(\mmssm)$ for $\tan\beta=3$ and
$\mmssm=1\tev$. In the lower shaded region, $M_{\stau_R}(\mmssm) >
M_1(\mmssm)$. In the upper shaded region, gauge coupling constants
become non-perturbative below the GUT scale. }
\label{fig:slM2}
\end{figure}

\begin{figure}[t]
\centerline{\epsfxsize=0.554\textwidth \epsfbox{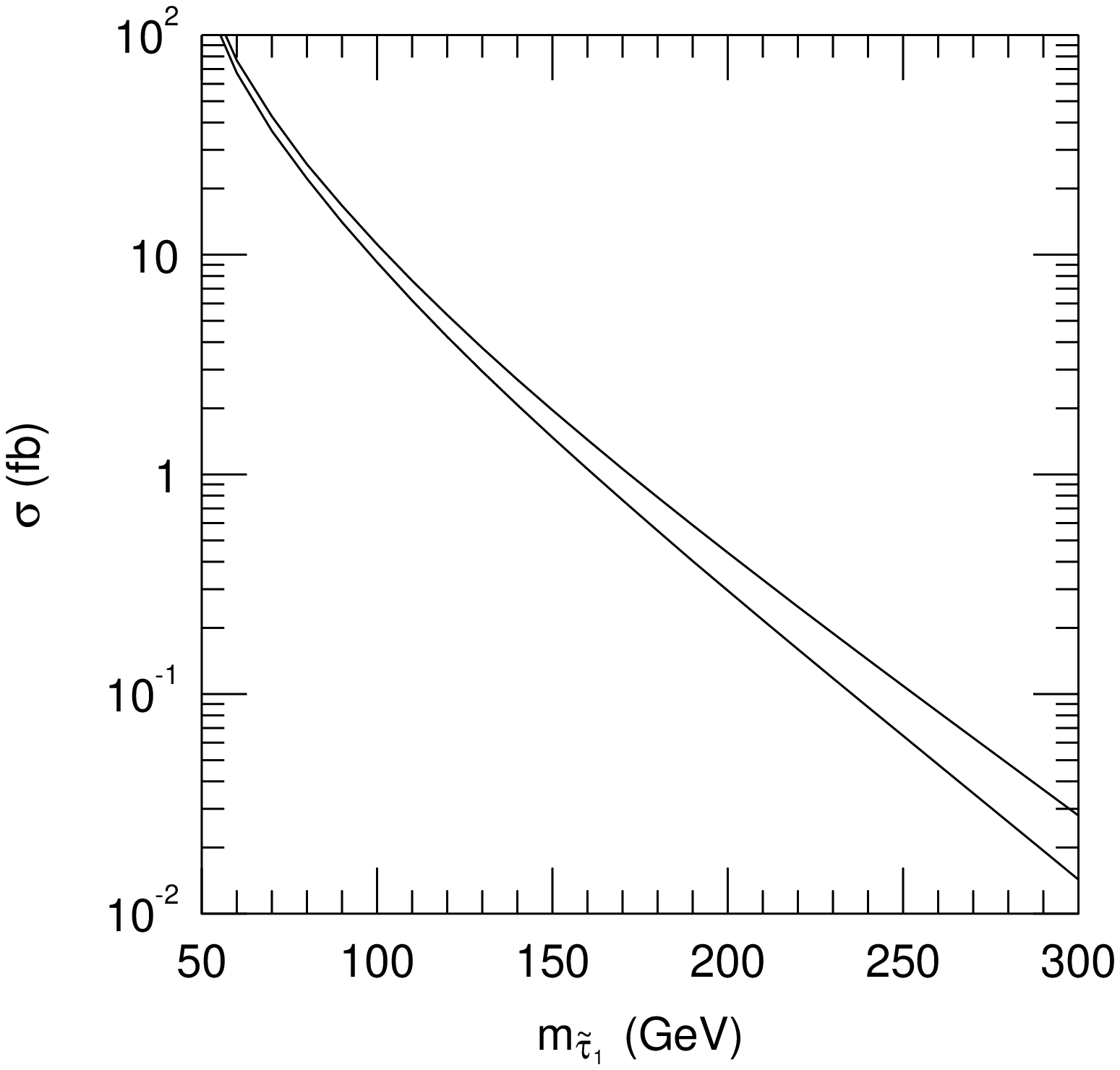} }
\vspace*{0.222in}
\caption{Cross sections for $p\bar{p} \to \gamma^*, Z^* \to 
\stauone \stauone^*$ for $\protect\sqrt{s} = 1.8$~TeV (lower) and 2~TeV
(upper), and $\stauone \approx \stau_R$. }
\label{fig:sl_prod}
\end{figure}

\begin{figure}[t]
\centerline{\epsfxsize=0.554\textwidth \epsfbox{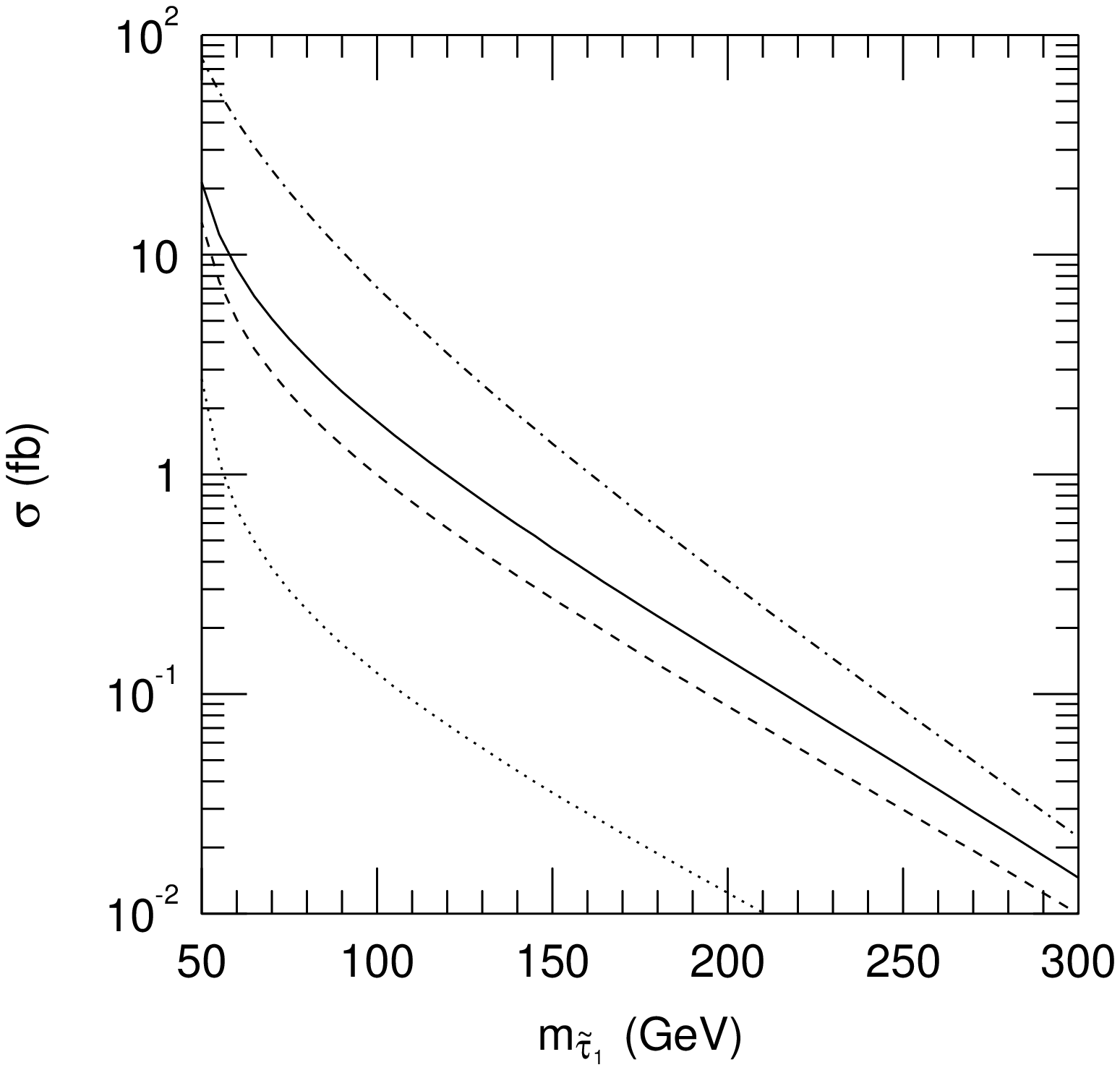} }
\vspace*{0.222in}
\caption{Cross sections for the production of at least one slow
$\stauone$ from $p\bar{p} \to \gamma^*, Z^* \to \stauone
\stauone^*$ at $\protect\sqrt{s} = 2 \tev$, where we have required
$|\eta|\leq 0.6$ for the slow $\stauone$. Contours correspond to
$\beta\gamma\leq 0.4$ (dotted), 0.7 (dashed), 0.85 (solid), and
$\infty$ (dot-dashed).}
\label{fig:ion_stau}
\end{figure}

\begin{figure}[t]
\centerline{\epsfxsize=0.554\textwidth \epsfbox{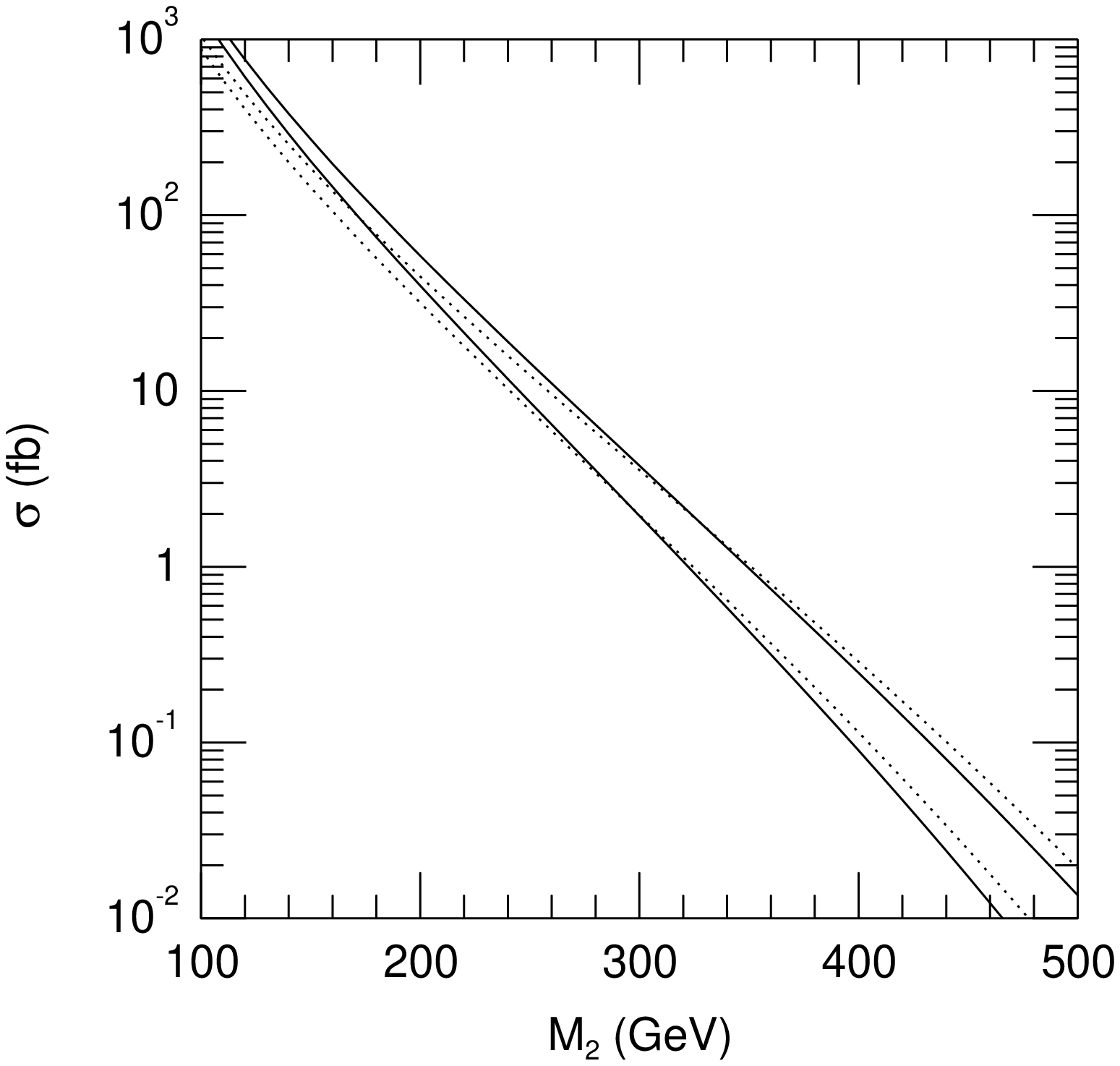}}
\vspace*{0.222in}
\caption{Cross sections for $p\bar{p} \to W^* \to \chi^\pm_1\chi^0_2$
(solid) and $p\bar{p} \to \gamma^*, Z^* \to \chi^\pm_1\chi^\mp_1$
(dotted), for $\protect\sqrt{s} = 1.8$~TeV (lower) and 2~TeV (upper).
Chargino and neutralino mixing effects are neglected. }
\label{fig:chi_prod}
\end{figure}

\begin{figure}[t]
\centerline{\epsfxsize=0.554\textwidth \epsfbox{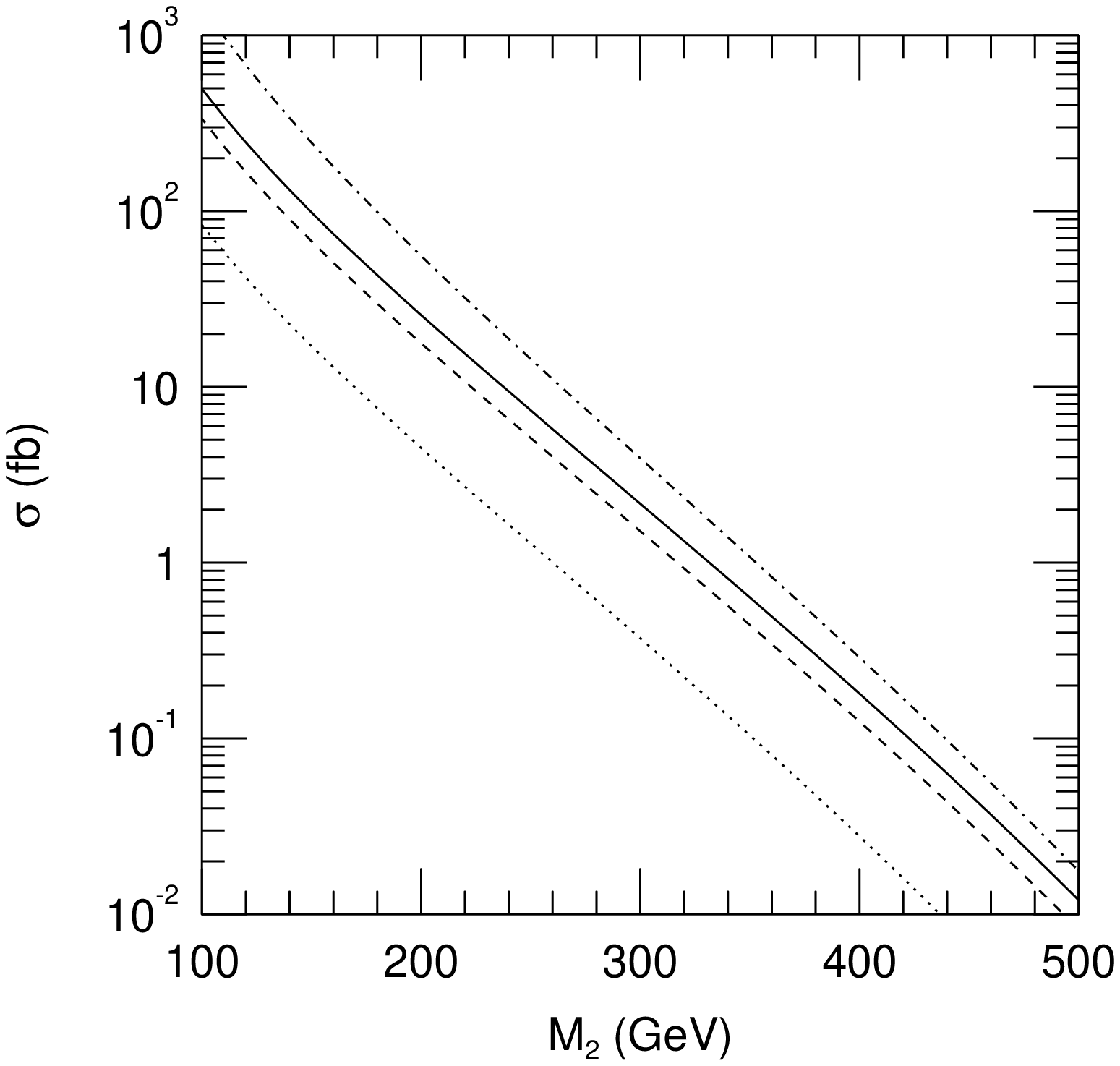} }
\vspace*{0.222in}
\caption{Cross sections for slow $\slepton_R$ production from chargino 
and neutralino production with $m_{\slepton_R} / M_2 = 0.5$, and
summed over the three slepton generations. Contours correspond to
$\beta\gamma\leq 0.4$ (dotted), 0.7 (dashed), 0.85 (solid), and
$\infty$ (dot-dashed). An efficiency of 50\% is included for the
requirement that the slow slepton have $|\eta|\leq 0.6$. }
\label{fig:ion_wino5}
\end{figure}

\begin{figure}[t]
\centerline{\epsfxsize=0.554\textwidth \epsfbox{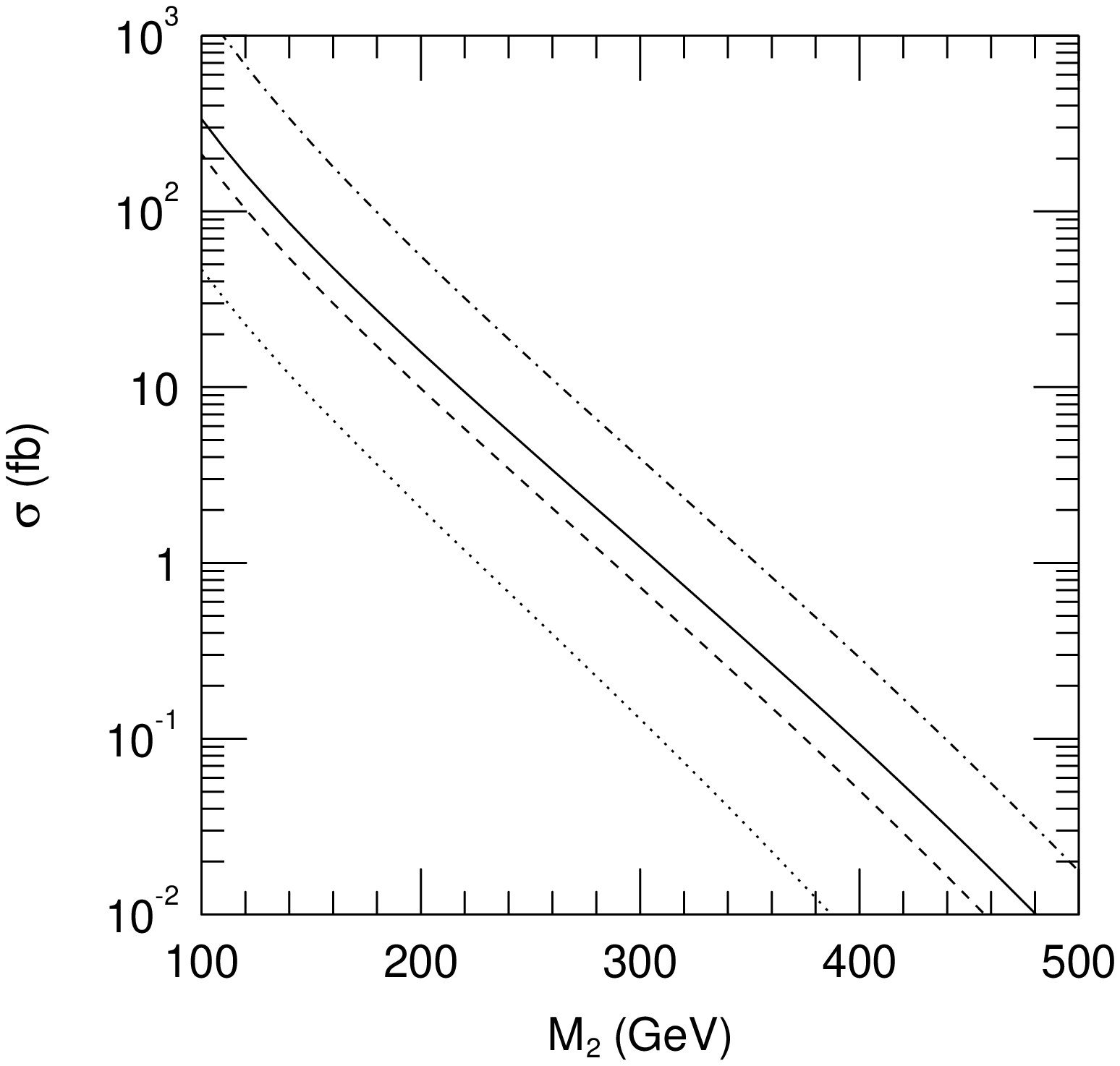} }
\vspace*{0.222in}
\caption{Same as in Fig.~\protect\ref{fig:ion_wino5}, but for 
$m_{\slepton_R}/M_2=0.4$.}
\label{fig:ion_wino4}
\end{figure}

\begin{figure}[t]
\centerline{\epsfxsize=0.554\textwidth \epsfbox{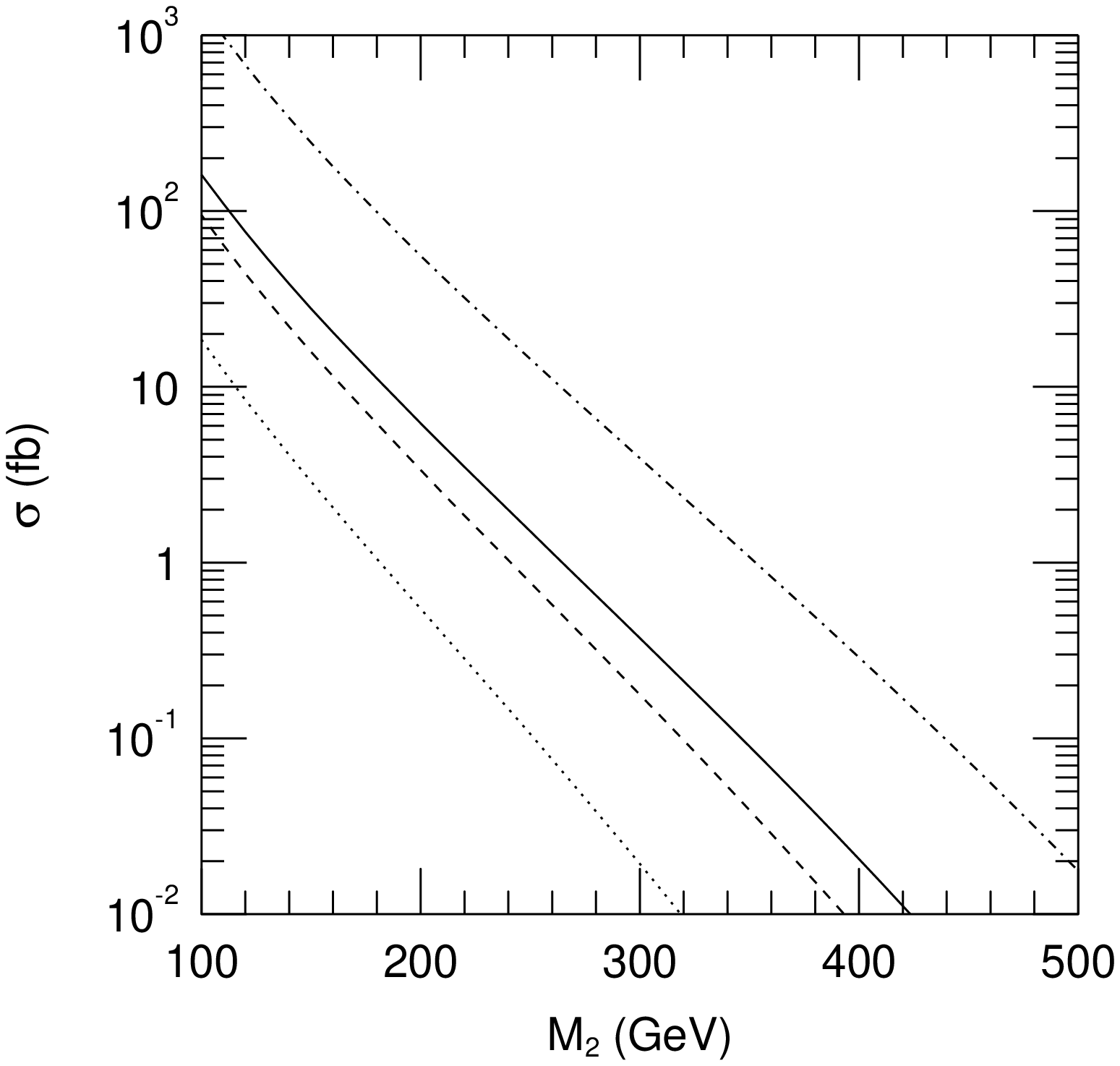} }
\vspace*{0.222in}
\caption{Same as in Fig.~\protect\ref{fig:ion_wino5}, but for 
$m_{\slepton_R}/M_2=0.3$.}
\label{fig:ion_wino3}
\end{figure}

\begin{figure}[t]
\centerline{\epsfxsize=0.554\textwidth \epsfbox{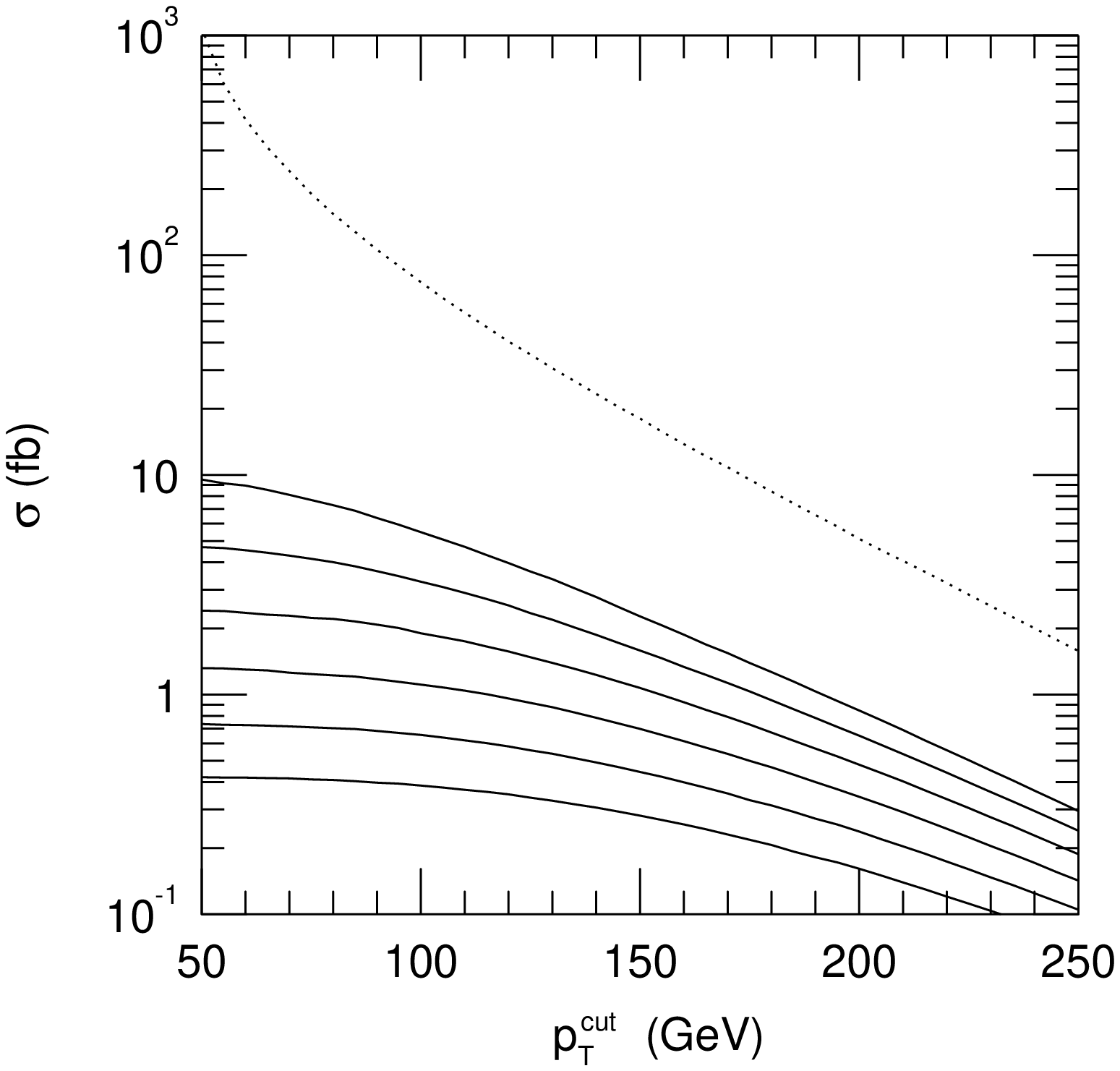}}
\vspace*{0.222in}
\caption{The cross section for $\stauone$ pair production at the 
Tevatron after requiring $p_T > \ptcut$ for each stau, as a function
of $\ptcut$. We require one $\stauone$ to have $|\eta| \le 0.6$ and
the other to have $|\eta| \le 1.0$, and we take
$\protect\sqrt{s}=2\tev$ and $m_{\stauone}=80$, 100, 120, 140, 160,
and 180~GeV (solid lines from above). The cross section for the
background $p\bar{p}\to\mu^+\mu^-$ is also shown (dotted). }
\label{fig:dimuon}
\end{figure}

\begin{figure}[t]
\centerline{\epsfxsize=0.554\textwidth \epsfbox{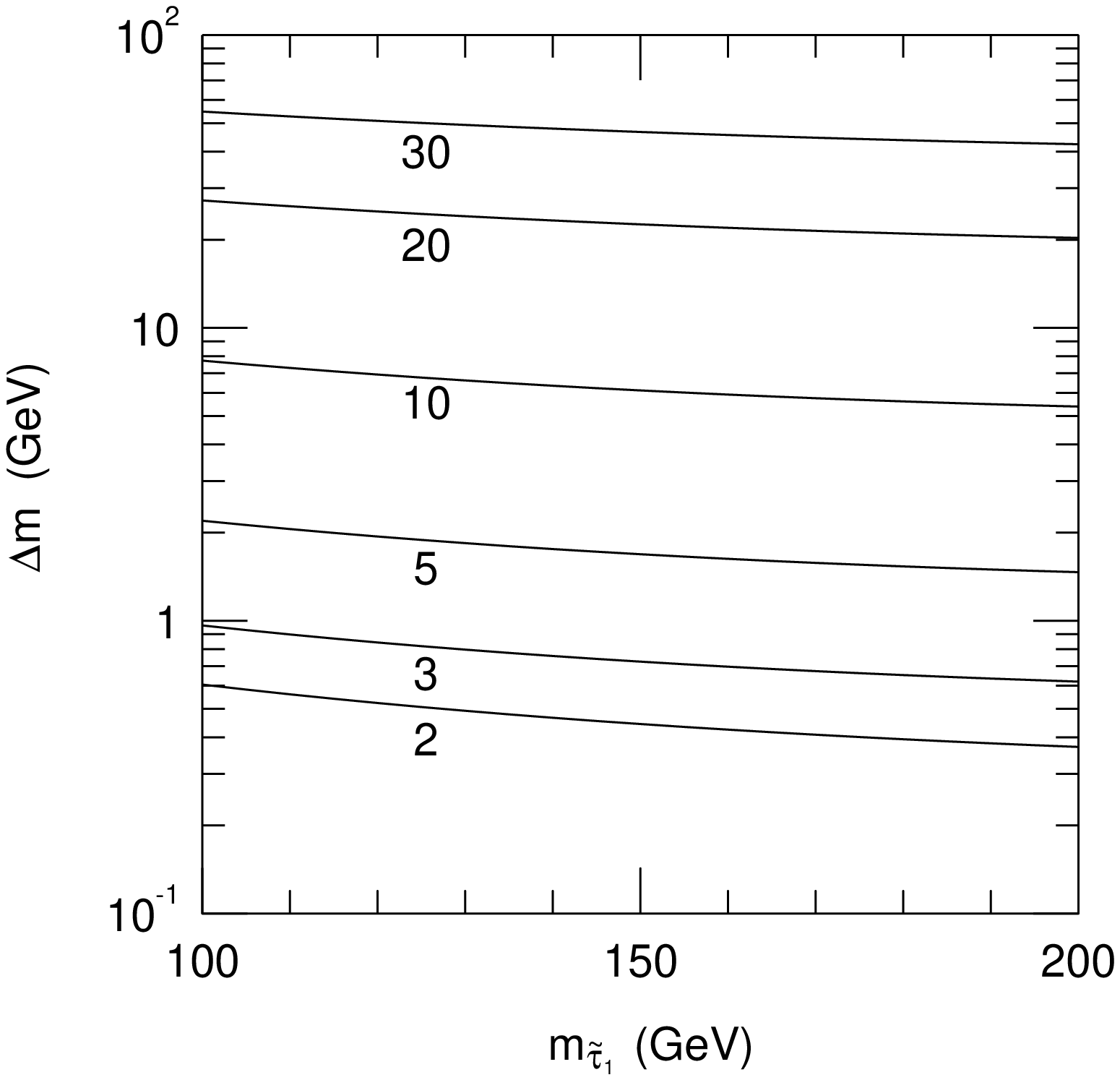}}
\vspace*{0.222in}
\caption{The mass splitting $\Delta m = m_{\selectron_R} -
m_{\stauone} \simeq m_{\smuon_R} - m_{\stauone}$ as a function of
$m_{\stauone}$ for fixed $\nn=3$, $\mmess=10^5 \gev$, $\mu > 0$, and
the $\tan\beta$ values shown. }
\label{fig:massdiff}
\end{figure}

\begin{figure}[t]
\centerline{\epsfxsize=0.554\textwidth \epsfbox{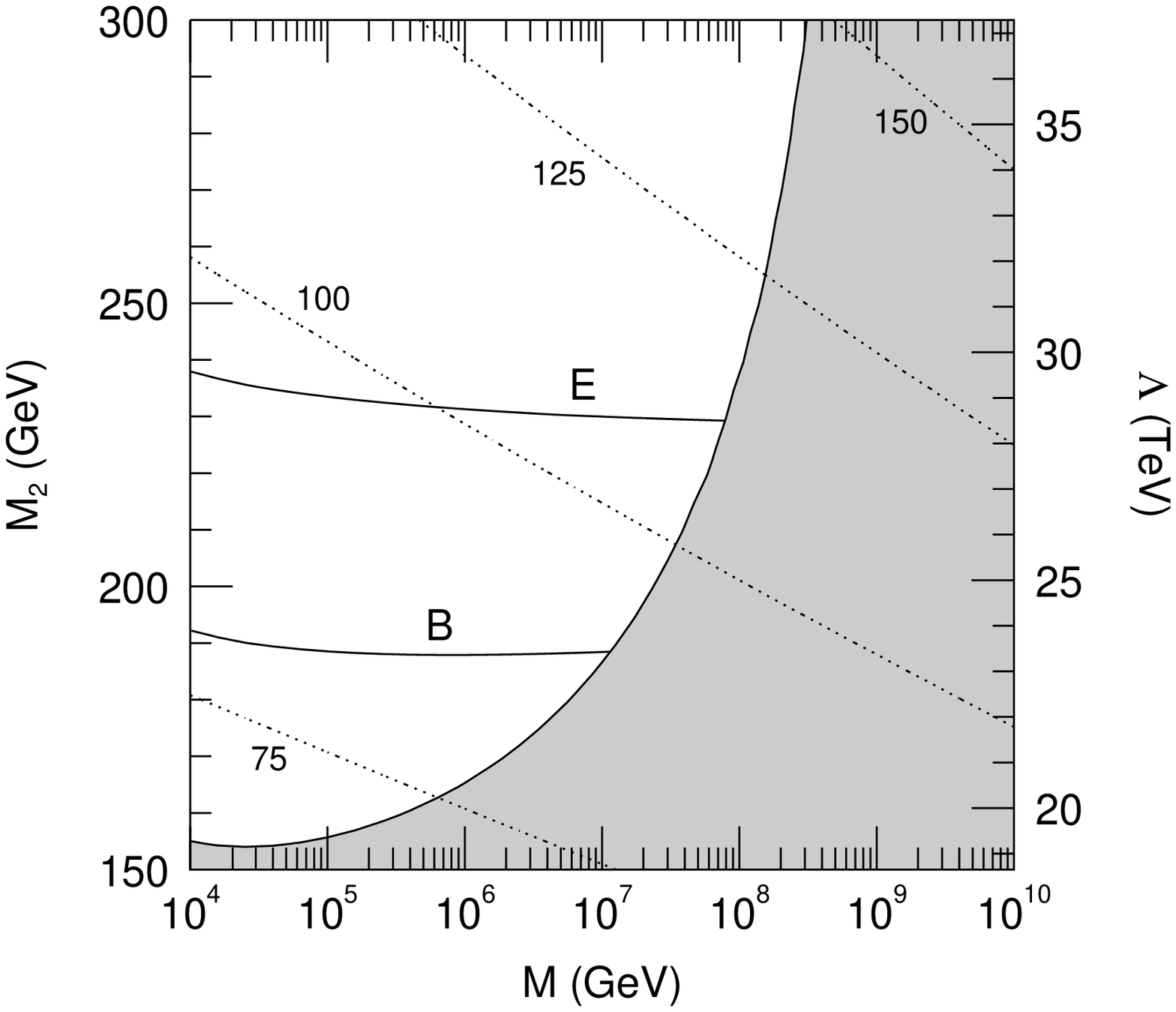}}
\vspace*{0.222in}
\caption{Summary plot for $\protect\sqrt{s} = 1.8\tev$ and
$L = 110\ipb$ (Run I), and fixed $\nn=3$, $\tan\beta=3$, and
$\mu>0$. Solid contours give the discovery reach for (B) highly
ionizing tracks from gaugino production, and (E) multi-lepton signals
from gaugino production.  Contours of constant $m_{\stauone}$ are
given by the dotted curves. In the shaded region, $\stauone$ is not
the NLSP.}
\label{fig:gm3_run1}
\end{figure}

\begin{figure}[t]
\centerline{\epsfxsize=0.554\textwidth \epsfbox{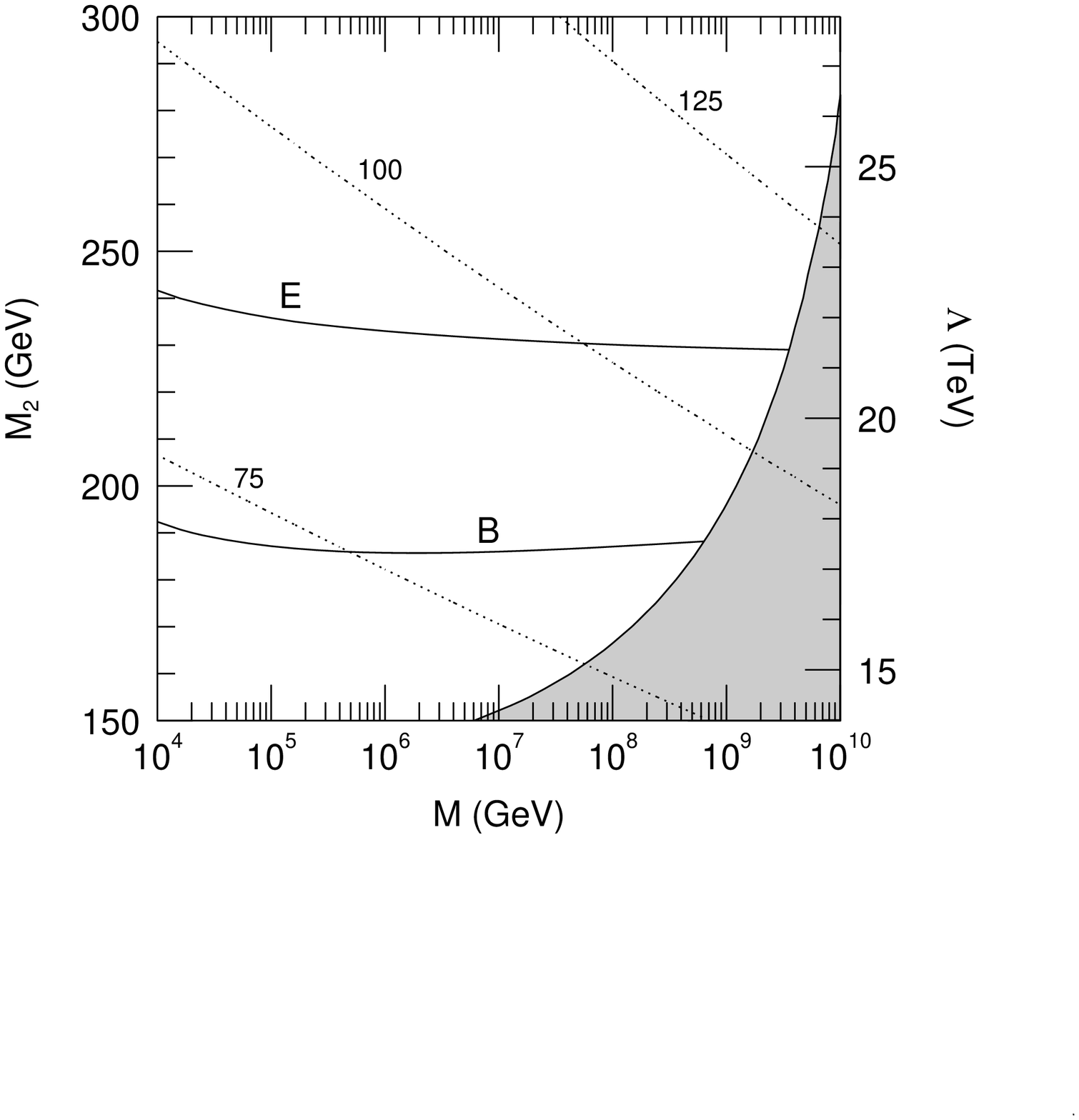}}
\vspace*{0.222in}
\caption{Same as in Fig.~\protect\ref{fig:gm3_run1}, but for $\nn=4$. }
\label{fig:gm4_run1}
\end{figure}

\begin{figure}[t]
\centerline{\epsfxsize=0.512\textwidth \epsfbox{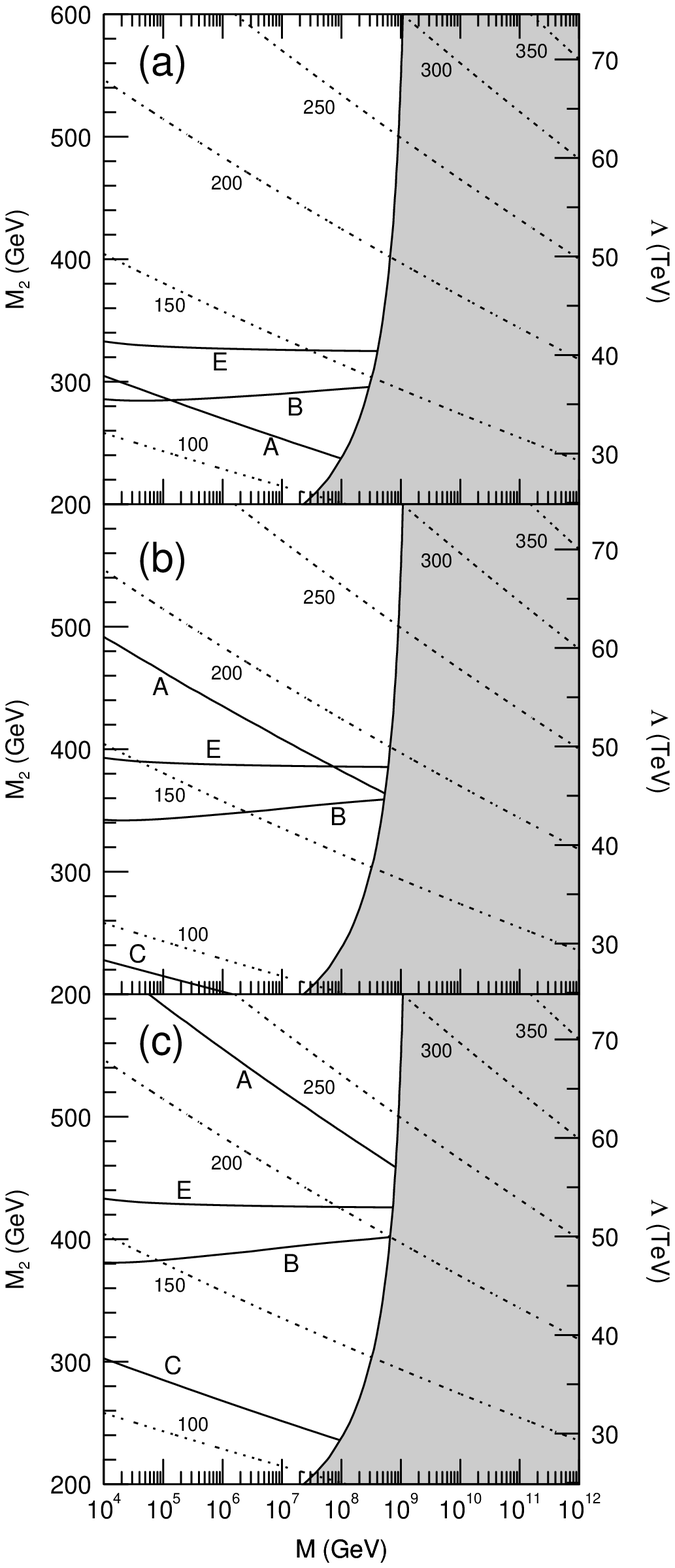}}
\vspace*{0.222in}
\caption{Summary plot for $\protect\sqrt{s} = 2\tev$ and integrated
luminosity (a) $2\ifb$, (b) $10\ifb$, and (c) $30\ifb$. We fix
$\nn=3$, $\tan\beta=3$, and $\mu>0$.  Solid contours give the
discovery reach for (A) highly ionizing tracks from slepton
production, (B) highly ionizing tracks from gaugino production, (C) a
dimuon excess from slepton production, and (E) multi-lepton signals
from gaugino production.  Five events are required for (A), (B), and
(E), and a $3\sigma$ excess is required for (C).  Contours of constant
$m_{\stauone}$ are given by the dotted curves. In the shaded region,
$\stauone$ is not the NLSP.}
\label{fig:gm3_tot}
\end{figure}

\begin{figure}[t]
\centerline{\epsfxsize=0.512\textwidth \epsfbox{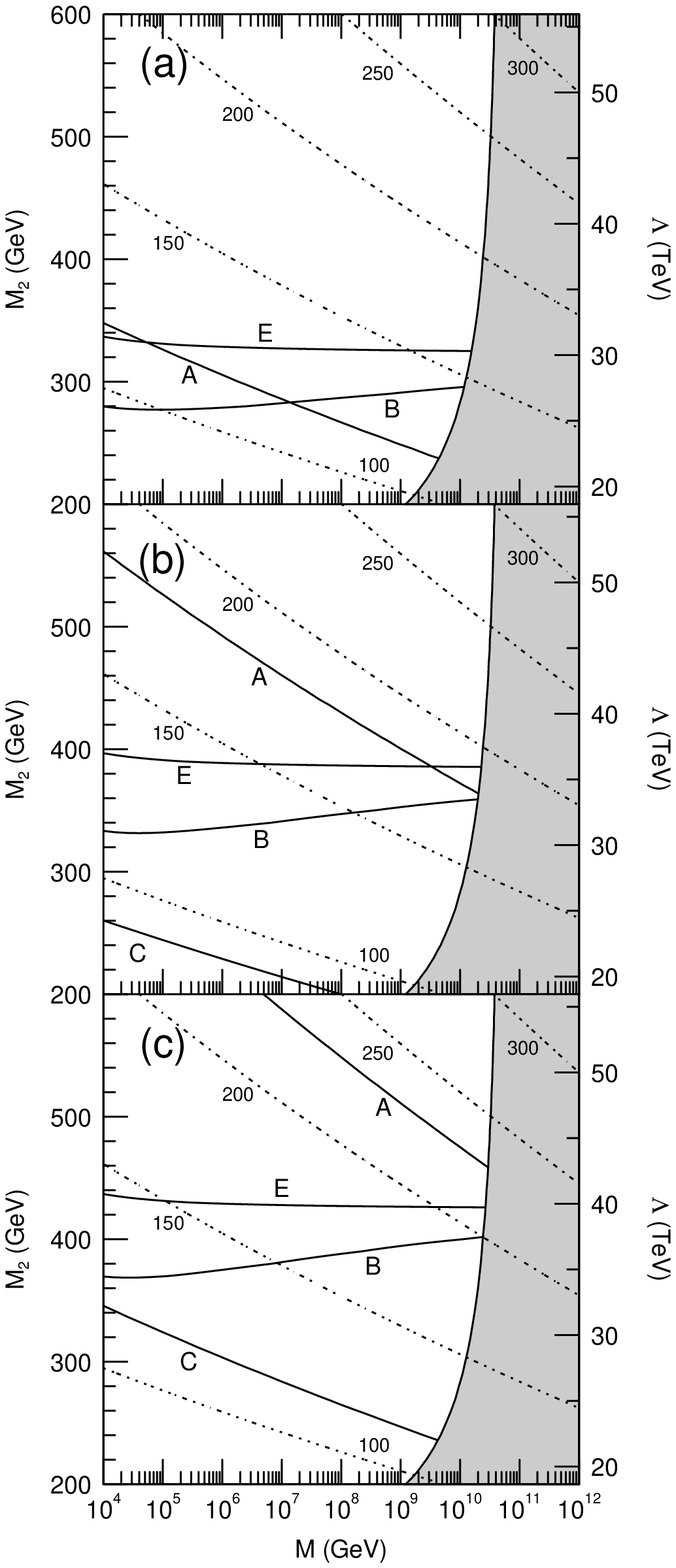}}
\vspace*{0.222in}
\caption{Same as in Fig.~\protect\ref{fig:gm3_tot}, but for $\nn=4$. }
\label{fig:gm4_tot}
\end{figure}

\end{document}